\documentclass[10pt]{article}
\pdfoutput=1
\usepackage{graphicx, verbatim, url, amssymb, amsmath, amsfonts, amsthm, latexsym, enumerate, float, subcaption}

\usepackage{booktabs}
\usepackage{cool}
\usepackage{graphicx}
\usepackage{multirow}
\usepackage[square]{natbib}
\usepackage[utf8]{inputenc}
\usepackage[english]{babel}

\begin{document}
	
\title{The structural role of the core literature in history\footnote{Submitted to \textit{Scientometrics}. This research is funded by the Swiss National Fund with grants 205121\_159961 and P1ELP2\_168489.}
}

\author{Giovanni Colavizza \\
	Digital Humanities Laboratory \\ 
	\'{E}cole Polytechnique F\'{e}d\'{e}rale de Lausanne -- Switzerland \\
	\url{giovanni.colavizza@epfl.ch} 
}

\date{}

\maketitle

\begin{abstract}
	The intellectual landscapes of the humanities are mostly uncharted territory. Little is known on the ways published research of humanist scholars defines areas of intellectual activity. An open question relates to the structural role of core literature: highly cited sources, naturally playing a disproportionate role in the definition of intellectual landscapes. We introduce four indicators in order to map the structural role played by core sources into connecting different areas of the intellectual landscape of citing publications (i.e. communities in the bibliographic coupling network). All indicators factor out the influence of degree distributions by internalizing a null configuration model. By considering several datasets focused on history, we show that two distinct structural actions are performed by the core literature: a global one, by connecting otherwise separated communities in the landscape, or a local one, by rising connectivity within communities. In our study, the global action is mainly performed by small sets of scholarly monographs, reference works and primary sources, while the rest of the core, and especially most journal articles, acts mostly locally.
	
\end{abstract}

\section{Introduction}
\label{sec:intro}

The traditional starting point to construct and analyse maps of science are citing publications and their cited sources. The literature humanists write comes in a variety of forms and with a multiplicity of intentions. \cite{hicks_four_2004} individuates four literatures in the social sciences, relevant for most disciplines in the humanities too: international journal articles, books, national literature (whose scope is local due to its topic and citation span, not necessarily its language) and non-scholarly publications. International journal articles are just the indexed tip of the iceberg. Books, albeit fewer in number, have a disproportionate impact in terms of received citations, and manifest specific citation behaviour patterns \citep{thompson_2002}. These publication typologies serve complementary purposes. In fact, if we follow \cite{nederhof_2006}, there are three target audiences for the social sciences and humanities: other scholars at the research frontier (assuming this group includes somewhat internationally recognized scholars); regional or national scholars (assuming scholars mostly dealing with national literature in the sense of \cite{hicks_four_2004}. Admittedly, the dichotomy international/frontier and, by implication, national or regional/not frontier, is probably too simplistic and unwarranted); and the non-scholarly public.

Taking the somewhat wider perspective of what can be cited by such literature, we find variety in abundance. A non-exhaustive list includes: primary sources (surely a black-box in itself), books, journal articles, conference proceedings and contributions in edited volumes, works of reference and edition of sources, databases and online resources, book reviews, plus all kind of writing for the general public such as other books, essays, newspaper or online articles and even blog posts. We should thus abandon any mono-dimensional view of humanities' scholars and the intellectual landscapes they inhabit \citep{watsonboone_information_1994,lariviere_canadian_2006}. More likely, several profiles of scholars in the humanities exist, each having a tendency for using a combination of the aforementioned typologies of sources and publications. Any effort to map the humanities must acknowledge their \textit{multidimensional intellectual organisation}, and the fact that using only a few of these sources will inevitably lead to a simplified view of their complexity.

We consider here two related questions. Firstly, can we say that a core literature exists in the humanities? In particular, we would like to know if a core literature exists for different citing publication typologies, and what it comprises. Secondly, what structural role does the core literature play with respect to the intellectual landscape defined by citing publications? More precisely, is the core literature spanning only specific locations, or does it connect far-apart areas of the landscape? Are different core sources, such as books and journal articles, behaving differently in this respect? In order to contribute towards an answer to these questions, we report on a case-study focused on history. We will consider results from three citation datasets. First, a monograph to book dataset on the history of Venice; secondly, all Web of Science (WoS)-indexed articles published in The Library, a renown journal in the area of the history of the book, with all WoS source and non-source items they cite; lastly, all historiography from WoS with all cited source items. In what follows we refer to books generally to indicate the variety of publications that take this form, including scholarly monographs, edited volumes, edited primary sources and reference works. 

In this article we contribute the following results:
\begin{enumerate}
	\item a general method to assess the structural role of cited publications into connecting the resulting bibliographic coupling network, via a set of complementary indicators.
	\item An analysis of the core literature in different datasets on history, showing that a core exists, and is mainly composed of books, where applicable.
	\item Evidence that the core literature can act both globally, or more often locally, and that only few sources, mostly books, provide global connectivity.
\end{enumerate}

This article is organised as follows. A first section discusses in some depth previous work on mapping the humanities, and contextual state of the art. We then present our methods and introduce the indicators. We then proceed to describe the datasets, present and discuss results, to then conclude.

\section{State of the Art}

\subsection{Science mapping}

Science can be conceptualized in a variety of ways, for example if can be viewed as a process of accumulation of new knowledge. Maps of science are attempts to localize and relate by relative positioning some entities of interest, such as publications, authors or journals, by way of some relations among them, for example citations \citep{borner_visual_2009}. Maps of science are especially helpful to uncover the cognitive structure, or intellectual organization of a discipline, in its constituents fields, sub-fields and topics of interest. The scale of analysis, the nature and variety of entities and relations, as well as the dynamics of the scientific process all play a role in this respect. If the mapping of science has been producing a growing number of contributions and increased understanding over time \citep{borner_visualizing_2003, boyack_mapping_2005, borner_atlas_2010}, the situation is less clear for the humanities. Often omitted from maps \citep{klavans_toward_2009}, ``the fine-structures of the humanities have been black-boxed and insufficiently unpacked; the available studies focused mainly on their positions relative to the social and natural sciences" \citep{leydesdorff_structure_2011}.

Usually, the research front can be mapped using bibliographic coupling, its intellectual base using co-citation networks \citep{persson_intellectual_1994}. The core literature, or highly cited sources, plays an important role in the intellectual base: ``the intellectual base is constituted by the core documents of a field; the documents that you should have read or cited, or the `classics' which you at least should be familiar with in order to be recognized as a member of the research community." \citep{hammarfelt_2011}. These consideration might not immediately apply to the humanities. The very notion of a research front has in this respect been questioned: ``the being-cited patterns [in some cases] do not indicate the provision of a knowledge base for new knowledge contributions at a research front, but may mean a source of cultural inspiration and influence. This would also explain the slower pace of `progress' in the humanities" \citep{leydesdorff_maps_2010}.

\subsection{Bibliometrics and the humanities}

The humanities are indeed considered to possess a set of characteristics which prevents to straightforwardly apply traditional bibliometric reasoning developed studying the sciences, and make it more challenging to acquire and use citation data to study their intellectual organisation and communication practices. Among them it is possible to find: the importance of the national and local dimensions, the variety of publication typologies, with a preference for monographs, the slower pace of theoretical development and citation accumulation, the richness of citation semantics (and syntax), the individual endeavour as the preferred way to organize research, the variety of sources and topics being investigated, and the resulting less focused and wider information retrieval behaviour \citep{garfield_is_1980,glanzel_bibliometric_1999,hicks_difficulty_1999,barrett_2005,nederhof_2006,huang_characteristics_2008,hellqvist_referencing_2009,linmans_why_2009,lin_citation_2013}. 

Consequently, it is more difficult to build comprehensive citation indexes in the humanities, a condition that hindered bibliometric research in this area \citep{ardanuy_sixty_2013}. This remains the case despite recent slow progresses, especially made by considering specific fields, important means of publication such as books, and new sources of data \citep{hammarfelt_2016}. These considerations in part motivate why, so far, ``the study of the intellectual structure within the humanities using citation analysis is as yet an underdeveloped area" \citep{hammarfelt_2011}.

\subsection{Empirical studies of intellectual landscapes in the humanities}

Nevertheless, we can still find a set of empirical studies which have been carried out in the humanities. We limit our attention to citation network analyses resulting in mapping efforts, a kind of bibliometric analysis seldom carried out for the humanities in the early decades of the discipline, in favour of more general descriptive citation studies \citep{herubel_citation_1994}.

Some studies considered specific journals or disciplines. In an early effort \cite{herubel_using_2001} analysed the French journal \textit{Annales} using the A\&HI, and assessing its international reach, as well as its capacity to rely on a broad array of literature from a variety of fields. \cite{leydesdorff_maps_2010} analysed two journals in the arts, Leonardo and the Arts Journal, using data from the A\&HI. The authors found that both journals cite mostly within the span of their original domain, but are cited widely outside of it. Differently, a small set of articles in the digital humanities was found to cite widely but being cited from a narrower community, resembling the sciences with respect to its ``being-cited patterns". The authors also add that: ``in the arts and humanities, one focuses on the tips of icebergs of possible references even more so than in the (social) sciences, since publication in the arts and humanities cannot be considered as an endogenous mechanism for generating and supporting a research front." \cite{coscia_exploring_2011} took the perspective of annotated bibliographies and their classification systems, a regular practice which aims at indexing the all new publications for specific disciplines in the humanities such as classics. By considering the \textit{Arch{\"a}ologische Bibliographie}, a bibliography for classical archaeology from 1956 to 2007, the authors proposed a way to explore both its classification system and, through it, the authors and publications of the bibliography at different levels of scale. They found that ``publications and authors in classical archaeology seem to specialize roughly on certain genres, governed by an either spatial, temporal, or a more generic conceptual perspective." The literature seems to organise itself by enriching and densifying a skeleton of clusters already in place since the 1950s in the classification co-occurrence network, either signalling that the literature has been incrementally growing in the well-defined fields of classical archaeology, or the conservative nature of the classification system of the \textit{Arch{\"a}ologische Bibliographie}. \cite{weingart_2015} analysed the fields of History and Philosophy of Science, relying on citation data from the A\&HI to both source and non-source items. Using bibliographic coupling and co-citation networks among journals and authors, the author showed how the two communities harbour a third community of authors at their border, who draw from both. Further in philosophy, \cite{ahlgren_bibliometric_2015} explored the ``subdomains" of free will and sorites, using co-citation maps at the level of authors, publications and journals, and terms co-occurrences, using A\&HI citation data. Interestingly, the authors found a mapping organized into fields of inquiry for free will, with important connections outside of philosophy proper (e.g. to neuroscience), and organized into smaller topics for sorites, consistent across different networks. More recently, \cite{colavizza_core_2017} considered the historiography on Venice using a novel monograph to book citation dataset. Results highlight the existence of a core literature, mainly composed of renown monographs, reference works and primary sources. This mostly quite aged core literature is the glue keeping the landscape united on the side of citing publications.

\cite{leydesdorff_structure_2011} provided for the first (and last) time an attempt to map all the humanities using the whole A\&HI index for the year 2008. Perhaps the most salient finding was a coherent set of twelve dimensions (latent factors) clearly organised in more or less proximal areas of research, among which we find classics, religion and archaeology; linguistics and the history and philosophy of science; literature and history; arts; music.

Different approaches were also considered. \cite{kreuzman_co-citation_2001} used author co-citation analysis in the fields of the philosophy of science and epistemology using A\&HI data. Multidimensional scaling was used in order to project the co-citation relations on a two-dimensional plane, broadly finding a division of authors according to the field or sub-fields and to the quantitative or qualitative approach. The perspective of author co-citation networks was also taken by \cite{hammarfelt_harvesting_2012} for the field of Swedish literary studies, finding a clear set of core, highly cited and influential authors, mainly emerging at an international level or from contemporary Swedish literature.

Yet another different perspective was taken by \cite{Zuccala_2015}, who ranked scholarly book publishers in historiography using citations to books from articles indexed in Scopus. The resulting map of publishers shows a strong polarity towards prestigious English or American publishers, with only some topical organization. A final aspect of the humanities, which has barely been explored, is mapping the use of primary sources. \cite{romanello_exploring_2016} considered data from \textit{L’Année Philologique}, an index of reviews of publications in the domain of Classics. The author was able to make preliminary efforts in the study classical authors, their works and even common quotes through their citation networks.

This overview of mapping efforts in the humanities highlights some commonalities:
\begin{itemize}
	\item The reliance on existing citation indexes, above all the A\&HI, with all its limitations, especially notable its bias for journal articles in English.
	\item The almost lack of general maps, but instead a focus on disciplines or fields of research.
	\item The presence of several attempts to counter the lack of data, e.g. by using non-source items, classification systems or novel datasets prepared with considerable effort.
	\item The still immature state of theoretical developments on the intellectual organization of the humanities.
\end{itemize}

\subsection{Books and core literature}

It should be clear by now that most of the efforts have not considered citations to books, and especially scholarly monographs among them. The main motivation for this absence is the lack of easily available data, which is often taken from the Arts \& Humanities Citation Index or Scopus. Their coverage, albeit improving over time \citep{Mingers_2015,waltman_review_2016}, is still not satisfying both for journals \citep{Mongeon_2015} and books \citep{Zuccala_2015}. Yet one of the main features of the humanities is their reliance on scholarly monographs. Monographs have been and still are the main publication channel in most disciplines in the humanities \citep{cullars_citation_1992,thompson_2002,knievel_2005,nederhof_2006,lariviere_place_2006,williams_role_2009,engels_changing_2012}: ``a monograph may tend to embody a more significant intellectual contribution and a synthesis of a larger body of research than a journal article" \citep{lindholm_1996}. 

As a consequence, the most cited literature in a field within the humanities should essentially comprise monographs \citep{hicks_difficulty_1999}, which would benefit in turn from the Matthew effect \citep{merton_matthew_1968} and become increasingly more popular: indeed some studies support this claim. For example, \cite{lindholm_1996} tracked journal article citations to a specific set of monographs, finding a group of core, highly cited sources in every discipline they considered. \cite{hammarfelt_2011} similarly found that 95\% of the 200 most cited references were monographs out of a set of journal articles in literature from the A\&HI. In a related way, \cite{hammarfelt_harvesting_2012} found a well-defined set of core authors in the intellectual base of Swedish literary studies. Furthermore, monographs are also the main cited publication typology of historians \citep{jones_characteristics_1972}, and are considered to be the most suitable publication typology to be used for bibliometrics studies dealing with the humanities \citep{chi_differing_2016}. Nevertheless other studies struggled to find a set of core sources. Such was the case for the seminal work on the history of technology by \cite{mccain_1987} or for a study on Nineteenth-century British and American literary studies \citep{thompson_2002}. More recently, \cite{nolen_2016} found ``a lack of any unifying core" in English literary studies, due to the diversity in the use of sources made by scholars. 

A possible motivation for the potential lack of a core, or of an easily detectable one, might be that the diversity of citation practices, even within the same discipline, is simply too broad to allow for a set of sources to emerge as a shared core \citep{thompson_2002}. It is known that the humanities present great variability in citation practices among their different disciplines \citep{knievel_2005}. Indeed \cite{heinzkill_2007} found that over 40\% of the monographs cited from a set of articles in English and American literature fall outside of the field as individuated by library classification. The humanities have also been found to undergo an increase in interdisciplinary citing in recent times \citep{leydesdorff_maps_2010,hammarfelt_2011}, which is also coupled by a growing international projection \citep{hicks_difficulty_1999,engels_changing_2012}. This might not help a core literature to emerge: ``a less demarcated discipline lacking a central core is heavily influenced by other research fields and therefore more interdisciplinary in referencing practices" \citep{hammarfelt_2016}. Furthermore, since publications in the humanities usually accumulate citations at a slower pace \citep{nederhof_2006,linmans_why_2009}, it follows that it is more difficult to study the most recent literature \citep{finkenstaedt_measuring_1990}, or the impact of individual scholars and institutions \citep{hammarfelt_2011}. A last characteristic shared by many studies is the small size of the datasets available, and the difficulty to reach sufficient coverage even within a small field.

\subsection{On history}

For historians too, a set of complementary publication typologies and publication levels exist. Historians privilege the monograph as the key means to publish the final result of a stream of research, which is carried out mostly individually \citep{knievel_2005}. Both non-serial and serial publications, despite taking quite some time to age, do become more rapidly obsolescent than primary sources \citep{jones_characteristics_1972}, that is to say the evidence on which scholars ground their work \citep{wiberleyjr_humanities_2009}. Primary sources in turn can be subject to transformations which influence their usage patterns, such as indexation and cataloguing in archives and libraries, publication in critical editions and digitization. The rapid rise in the number of online available sources, both primary and secondary, might in fact be the most important change in recent historiography, despite the largely passive attitude of historians in this respect \citep{hitchcock_confronting_2013}. Another under-appreciated source for historians are reference works, such as dictionaries, catalogues and repertories. Eventually, historians are particularly sensible to two forms of localism: linguistic and geographical \citep{kolasa_specific_2012}, mainly due to the language and location of their sources. To summarize, historians use a wide array of materials, resulting in a likely large fraction of rarely cited items, yet we would expect few of these to be highly cited core sources, with a longer than average life-span, some with an indefinite life span too. In principle and in practice, there is no reason to think that core sources should only be books, even if most, likely are.

The literature is uniform in highlighting the importance of books in the publication practices in the humanities, but contradictory with respect to the presence of a core set of sources. Claims on the variety of sources used by humanists, their interdisciplinary citation practices and the blurred boundaries that separate their communities do not completely match with the presence of a well-individuated core literature. Furthermore, should a core literature exist, little is known about its structural role: are core sources emerging within a discipline or a field, or even a sub-field and lower levels? Do they instead bridge different areas of the landscape they emerge from?

\section{Method}\label{sec:method}

Given a set of citing publications, representing an area of research of interest, its core literature can be defined in a variety of ways. We consider here as core the most cited publications in the given dataset. Such core literature can then be analysed with respect to the bibliographic coupling network of the citing publications, which defines the intellectual landscape of the area of research. More specifically, with respect to a partition of such network into communities. Given that this network's structure necessarily relies to a considerable degree on the core literature, our goal is to qualify different roles that core sources can assume with respect to the communities of the bibliographic network. Four indicators, defined for a core source $c$, are introduced:
\begin{itemize}
	\item \textit{Within indicator} $a_c$: captures the importance of the source $c$ to connect citing publications within the same communities.
	\item \textit{Between indicator} $b_c$: captures the importance of the source $c$ to connect citing publications across communities.
	\item \textit{Topicality indicator} $c_c$: captures the relative importance of the source $c$ to connect citing publications within a specific community or within several communities. Topicality quantifies how focused the action of a source $c$ is within a specific community.
	\item \textit{Bridging indicator} $d_c$: captures the relative importance of the source $c$ to connect citing publications between a specific pair of communities or between several pairs. Bridging quantifies how focused the action of a source $c$ is across a specific pair of communities.
\end{itemize}
The four indicators capture different aspects of the role of the core literature with respect to relations between citing publications. They indicate how important the core source is to connect communities internally (within) or among each other (between), and how focused this action is (most influence within one community or between a pair of communities).

More formally we start with the following setting. Take $D = (V_D,E_D)$ the directed citation network of the set of publications under consideration, where vertices $v \in V_D$ are both citing publications and cited sources, and $e(v_1,v_2) \in E_D$ are directed edges between such vertices. A source can be both citing and cited, thus $D$ is not in principle acyclic. Take the projection of $D$ onto citing publications $B = (V_B,E_B)$: the weighted bibliographic coupling network. $B$ can also be represented by its square and symmetric adjacency weighted matrix $W$. For simplicity, without loss of generality, we consider raw weights of one for each reference in common between any two citing publications. Take $L$, a partition of $B$ into communities, where every vertex $v \in V_B$ is assigned to a unique community. Lastly, take a set of core sources $C \subseteq V_D$. Core sources can be individuated in a variety of ways, for example by taking a certain top quantile of the in-degree distribution (number of received citations) of $D$. To be sure, any cited source in $V_D$ can be considered for analysis, core sources being just an interesting subset of them.

All indicators are based on the idea of considering the contribution of a source in the core $C$ to the weight of the edges in $B$, the bibliographic coupling network, taking into consideration its partition into communities. Consider the function:
$$
cit(i,j,c) = \begin{cases}
1       & \quad \text{if } \exists e(i,c) \in E_D \wedge \exists  e(j,c) \in E_D \\
0  & \quad \text{otherwise}\\
\end{cases}
$$
That is to say, if both $i$ and $j$ cite $c$ in $D$, $cit(i,j,c)$ returns 1, which is the weight contributed by $c$ in the edge between $i$ and $j$ in $B$. This function assumes raw weights were used to construct the bibliographic coupling network. That is to say, $W_{i,j} = \sum_{v \in V_D} cit(i,j,v)$. Other weighting schemes might be used, such as fractional counting \citep{perianes-rodriguez_constructing_2016}, and then $cit$ shall be modified accordingly.

We can now proceed to establish a preliminary version of our indicators, defined for every $c \in C$ as follows:
$$
\begin{array}{lcl}
\alpha_{c} &=& \sum_{i,j \in V_B} cit(i,j,c) \delta(l_i,l_j) \label{eqn1} \\
\\
\beta_{c} &=& \sum_{i,j \in V_B} cit(i,j,c) (1-\delta(l_i,l_j)) \label{eqn2} \\
\\
\gamma_{c} &=& \max_l \alpha_{c}(l) \label{eqn3} \\
\\
\delta_{c} &=& \max_l \beta_{c}(l) \label{eqn4} \\
\\
\end{array}
$$

Where $l_i$ is the community to which $i$ is assigned according to partition $L$, and $\delta(l_i,l_j) = 1$ if $l_i = l_j$, $0$ otherwise. Note that $a_c + b_c = \sum_{i,j} cit(i, j, c)$, the total edge weight contributed by $c$ in $B$.

Yet the degree distribution of citation networks is normally skewed, thus the core has by definition a disproportionate role in the structure of the bibliographic coupling network. The problem with these indicators is that they do not account for the obvious effect of the in-degree of core sources. We would like, instead, to be able to compare different core sources, and core sources from different datasets, irrespective of the underlying degree distribution. As a consequence, we need a null model to compare against. The obvious choice is the \textit{configuration model} (cf. \cite{newman_2010,barabasi_network_2016}). In a directed setting, having the list of vertex pairs in two arrays (citing and cited respectively) of equal length, an instantiation of the configuration model consists of randomly permuting one of the two arrays, to produce a random network with the same degree distribution as the original one. A minor adaptation is the need to `simplify' the so-created random network by removing eventual self-loops and multi-edges (low-probability events in themselves). Such network can serve as a null model to test some properties of the original network, disregarding the effect of its degree distribution. It is good practice to produce several instantiations of such configuration model and average out the desired statistics.

In our case, we take $N$ instantiations of the configuration model of the directed network $D$, each time construct a new bibliographic coupling network and calculate as follows:

$$
\begin{array}{lcl}
\chi_{c} &=& \sum_{n}^N \sum_{i,j \in V_{B}} cit^n(i,j,c) \delta(l_i,l_j) = \sum_{n}^N \chi_{c}^n \\
\\
\phi_{c} &=& \sum_{n}^N \sum_{i,j \in V_{B}} cit^n(i,j,c) (1-\delta(l_i,l_j)) = \sum_{n}^N \phi_{c}^n \\
\\
\psi_{c} &=& \sum_{n}^N \max_l \chi_{c}^n(l) \\
\\
\omega_{c} &=& \sum_{n}^N \max_l \phi_{c}^n(l) \\
\\
\end{array}
$$

Where $cit^n$ considers edges in the $n$th instantiation of the configuration model. Note that we keep the same partition $L$ at all times.

The final indicators are:

$$
\begin{array}{lcl}
a_{c} &=& \frac{\alpha_{c}}{\alpha_{c} + \beta_{c}} - \frac{\chi_{c}}{\chi_{c} + \phi_{c}} \label{eqn9}\\
\\
b_{c} &=& \frac{\beta_{c}}{\alpha_{c} + \beta_{c}} - \frac{\phi_{c}}{\chi_{c} + \phi_{c}} = - a_{c} \label{eqn10} \\
\\
c_{c} &=& \frac{\gamma_{c}}{a_{c}} - \frac{\psi_{c}}{\chi_{c}} \label{eqn11}\\
\\
d_{c} &=& \frac{\delta_{c}}{b_{c}} - \frac{\omega_{c}}{\phi_{c}} \label{eqn12} \\
\\
\end{array}
$$

The behaviour of the indicators is as follows: $a_c$ is positive if the core source is more important than its degree would justify in connecting nodes within the same communities, $b_c$ if across communities. The more positive $c_c$ is, the more the action of $c$ in connecting nodes happens within the same community, the more positive $d_c$ is, the more the action happens between the same pair of communities, irrespective of the effect of the in-degree of $c$. Note that $a_c$ is in general positively correlated with the modularity of the partition $L$ \citep{newman_finding_2004}, and $b_c$ negatively so. As a consequence, any value of these two indicators is related to a given partition of the nodes. If the partition in use is the result of a stochastic modularity maximization procedure, all indicators should thus be averaged over multiple partitions. 

The relation of $a_c$ (and $b_c$) to the modularity of the partition $L$ is of interest. Modularity measures the density of the links within communities, as established by $L$, by comparing it to a random null model. In our setting, the (weighted) modularity $Q$ of $L$ is defined as:

$$
Q = \frac{1}{2w} \sum_{i,j \in V_B} \Bigg[ W_{i,j} - \frac{W_{i,*}W_{j,*}}{2w} \Bigg] \delta(l_i,l_j)
$$

Where $w$ is the sum of the weights of all edges in $W$, the weighted adjacency matrix of $B$, and $W_{i,*} = \sum_{j} W_{i,j}$, the weighted degree of vertex $i$. If we take the perspective of $c$, our core source of interest, we can construct a bibliographic coupling network $B^c = (V_B,E^c_B)$, and related $W^c$, only on the basis of the edges established by coupling citations to $c$. Note that the adjacency matrix $W^c$ is binary at this point. In this setting, an alternative definition to the within indicator (and the between, similarly) can be based on modularity as follows:

$$
a^*_c = \frac{1}{2w^c} \sum_{i,j \in V_B, \exists e(i,j) \in E^c_B} \Bigg[ W^c_{i,j} - \frac{W_{i,*}W_{j,*}}{2w} \Bigg] \delta(l_i,l_j)
$$

That is the modularity of $L$ considered only on the weights contributed by $c$, whose total sum is $w^c$. The main difference between $a_c$ and $a^*_c$ rests in the use of different null models, calculated over different networks: the configuration model establishes random multigraphs with a given degree sequence, here over $D$, the directed network; the null model used in modularity, called the Chung-Lu model, establishes random simple graphs with a given degree distribution, here over $B$, the bibliographic coupling network. The modularity $Q$ of $L$ given above is an aggregated function of this last alternative indicator, therefore the general distribution of both the within and between indicators are influenced by it. Nevertheless, individual core sources can behave in a variety of ways under this general setting.

\section{Datasets}\label{sec:data}

We use three datasets, motivated by the desire to consider the role of different core literature and citing publications with respect to their publication typology. The main difference cast here relates to the distinction between monographs and journal articles, both as citing and cited sources. In order to attempt such a comparison, we had to consider quite different datasets: a) a first dataset of monograph to book citations in the sub-field of the history of Venice, largely not indexed elsewhere; b) a dataset of article to both book and article citations, extracted from a specific journal in the sub-field of the history of the book, called The Library. We considered its WoS-indexed articles and what they cite, either source and non-source in WoS; c) a third dataset of article to article citations, from all WoS subject `History', limiting cited sources to what is indexed in WoS. These datasets should allow us to discuss sets of core literature composed by books cited by monographs, articles and books cited by articles, and articles cited by articles, yet it must be stressed that they are compared not as equals, but as a means to explore the same phenomenon from different angles. A summary of the three datasets is given in Table \ref{tab:general_stats}. There is a disparity in terms of size and coverage among these datasets, largely due to data availability. This especially entails the fact that dataset three, the largest one, should be considered somewhat apart of the other two, as will be discussed in what follows.

We identify the core literature in each dataset by taking all sources in the top $99.5$ percentile of the in-degree distribution (number of received citations) of every directed graph. All in-degree distributions are skewed (omitted here) and highlight a few, highly cited sources.

\begin{table}[H]
	\centering
	\caption{Summary statistics for the three datasets under consideration. The threshold of the number of received citations at the 99.5 quantile, used to establish the set of core sources, is given in the last row.}
	\label{tab:general_stats}
	\resizebox{\columnwidth}{!}{%
		\begin{tabular}{cccc}
			\textbf{Statistic/Dataset}               & \textbf{Monographs History Venice} & \textbf{The Library 1981-2016} & \textbf{All History 2005-2015} \\
			\textbf{Citing typology}                 & Monographs                         & Journal articles               & Journal articles                      \\
			\textbf{Cited typology}                  & Books                         & Books and articles               & Journal articles                      \\
			\textbf{\# citing publications}   & 673                                & 479                            & 36709                                 \\
			\textbf{\# cited sources}         & 36922                              & 11237                          & 101777                                \\
			\textbf{\# edges}                        & 68525                              & 13176                          & 159610                                \\
			\textbf{\# core sources (99.5 quantile)} & 129 (22)                           & 65 (6)                         & 776 (9)                              
		\end{tabular}
	}
\end{table}

In order to find the partitions of the bibliographic coupling networks we use a modularity maximization approach \citep{newman_finding_2004}, in the popular Louvain implementation \citep{blondel_fast_2008}. Despite its shortcomings, this method produces high quality results and is widely known in both the networks \citep{fortunato_community_2016} and bibliometrics communities \citep{subelj_clustering_2016}. All indicators were calculated averaging results from ten possible partitions made using the Louvain algorithm with default resolution parameter at one, and for each one a hundred instantiation of the configuration model were averaged.\footnote{Analyses relied on igraph [0.7.1] \citep{igraph} and Vincent Traag's community detection library [0.5.3] available at \url{https://github.com/vtraag/louvain-igraph}.} Modularity maximization tends to produce larger communities with larger datasets, when using the same resolution parameter. This is important to keep in mind for dataset three, which is larger in size than datasets one and two.

\subsection{Monographs of the historians on Venice}\label{sec:venice}

The first dataset considers the specific sub-field of the history of Venice. This dataset comprises relatively recent monographs in a variety of languages, selected through library resources such as catalogues and shelving strategies. The procedure followed to extract their citation data is detailed elsewhere \citep{colavizza_ijdl_2017}. The dataset is freely available online in multiple formats \citep{dataset}. This dataset comprises 673 citing monographs, which cite 36922 books in turn, with 68525 unique citations.

The dataset of citing monographs comprises works from different communities: medieval, early modern and modern history, art history and history of architecture, plus a variety of specialities therein, such as economic or gender history. The core literature of this dataset is mainly composed of primary sources, works of reference and renown scholarly monographs \citep{colavizza_core_2017}. Primary sources are often edited documents (e.g. the diaries of Marin Sanudo) or early printed works. Reference works such as repertories, inventories and dictionaries, often product of the local historians of the XIXth century, are still largely in use today. Lastly, scholarly monographs of lasting importance include some works from the XIXth century, as well as more recent literature published since the 1950s.

\subsection{The Library: articles of the historians of the book}\label{sec:library}

Our second dataset considers a different field of history: the history of the book, and a different publication typology: journal articles. We consider one of the most renown journals in this context: `The Library: Transactions of the Bibliographical Society' of London. Historians of the book extensively rely on resources such as catalogues and repertories for their work, they are organised into quite specialised communities with strong bonds with other fields such as library and information science, literature and philology.

All the indexed research articles in WoS are considered, from 1981 to 2016 included, for a total of 491 articles. We consider both source and non-source items by exporting all references from the WoS interface. Exported articles were processed using the Sci2 tool [1.2 beta] \citep{sci_tool}, in order to extract the directed citation network. Sci2 allows to detect duplicate nodes (i.e. references pointing to the same item) by comparing all references using the Jaro-Winkler measure.\footnote{See \url{http://wiki.cns.iu.edu/display/CISHELL/Detect+Duplicate+Nodes}, accessed May 2017.} Groups of references similar above a certain threshold are retained as candidates for merging. This method is far from perfect, but allows to create a set of grouped references to be manually checked for refinement. We proceeded as follows: first, all references with no author were removed from the dataset, as too problematic to disambiguate. This is the case, for example, for references to newspaper articles. Secondly, all paginations were removed, given that they often pointed to the specific location which was cited (as it is practice in the humanities) instead of the page interval of the cited article (as in the sciences). The Sci2 tool was then used to detect groups of references to be merged, with a threshold of 0.84 on the Jaro-Winkler measure, established empirically by finding the threshold, rounded at two digits, which would yield a precision of less than 0.5 in the 100 pairs of references to be merged with a similarity just below that threshold. On this dataset, a threshold of 0.84 had a precision so calculated of 0.41. Note that we left inevitably out some references as not grouped, but the number of false negatives decreased rapidly thereafter. Lastly, all retained groups of references to be merged were manually checked and cleaned. During cleaning, multiple editions of the same work were considered as one.
The result is that 479 articles cite 11237 unique items. Some citing articles are removed because they did not possess extracted references (9) or they were merged (2). The number of references in the original dataset is 14412, the number of citations after clean-up is 13180.

The core literature of this dataset is mainly composed of primary sources, works of reference and seminal monographs. Examples include the overly important Shakespeare, the records of the Stationers Company and early printing manuals such as Moxon's Mechanick Exercises; catalogues (e.g. the English Short Title Catalogue), dictionaries or reference works (e.g. Plomer's Dictionary of Booksellers and Printers); renown monographs (such as Gaskell's New Introduction to Bibliography). Only 13 of the 100 most cited sources are journal articles. The core literature resembles the one from Venice in its assortment, with quite more emphasis on reference works: the cornerstone of studies in bibliography and the history of the book.

\begin{figure}[h!]
	\centering\includegraphics[width=\columnwidth]{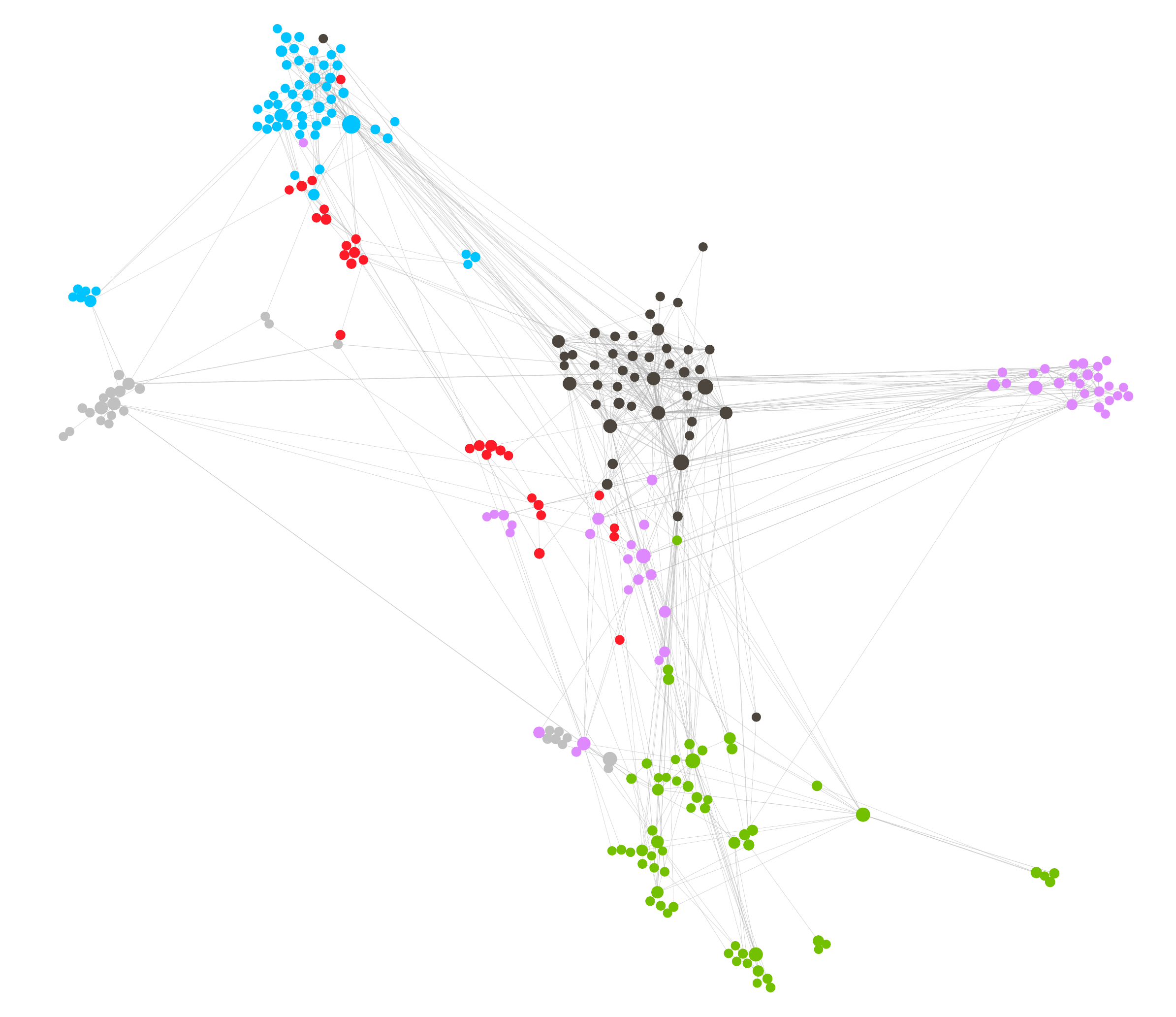}
	\caption{The communities of the Library dataset, using the Louvain method with configuration identical to experiments. This network has been trimmed from edges of weight less than 2, as a consequence nearly 60\% nodes are visible. The communities are: cyan -- early English printing; pink -- Shakespeare; green -- English literature XVI-XVII century; grey -- Renaissance book production in the European continent; dark grey -- English book production and commerce in early modern times; red -- Libraries and collections. The size of the nodes is proportional to their betweenness centrality. This visualization was made with Gephi 0.9.1 \citep{gephi}, using Force Atlas 2 with default parameters but for LinLog, dissuade hubs and prevent overlap modes active, scaling 2.0 and edge influence 1.5 \citep{jacomy_forceatlas2_2014}.}
	\label{fig:library_comm}
\end{figure}

The main intellectual communities of this dataset are shown in Figure \ref{fig:library_comm}. The strong focus on English studies, and the interconnection of the history of the book and literature clearly emerges, in particular for Shakespeare studies. Topics from continental book history seem marginal instead. Yet the community publishing in The Library appears well organised in specific areas of activity. 

\subsection{All of history in the Web of Science}\label{sec:wos}

The last datasets comprises all articles indexed in WoS under the WoS subject category of `History', published from 2005 to 2015 included. No other publication typology besides research articles was considered. With respect to their citations, everything that was indexed in WoS is retained, also outside of this specific subject category. Citations were taken from the CWTS databases \citep{OSE16}. This dataset comprises 36709 citing articles and 101777 cited articles, with 159610 citations among them.

The core literature of this dataset is composed mainly of seminal articles which delivered novel methods or arguments of enduring importance for a broad area of historical studies. Examples include gender history (Scott, 1986, \textit{Gender, a useful category for historical analysis}), politicization of the past (Hall, 2005, \textit{The long civil rights movement and the political uses of the past}), comparative history of development (Acemoglu, 2001, \textit{The colonial origins of comparative development: An empirical investigation}) and cultures (Subrahnanyam, 1997, \textit{Connected histories: Notes towards a reconfiguration of early modern Eurasia}), political history (Elliott, 1992, \textit{A Europe of composite monarchies}).

\begin{figure}[h!]
	\centering\includegraphics[width=\columnwidth]{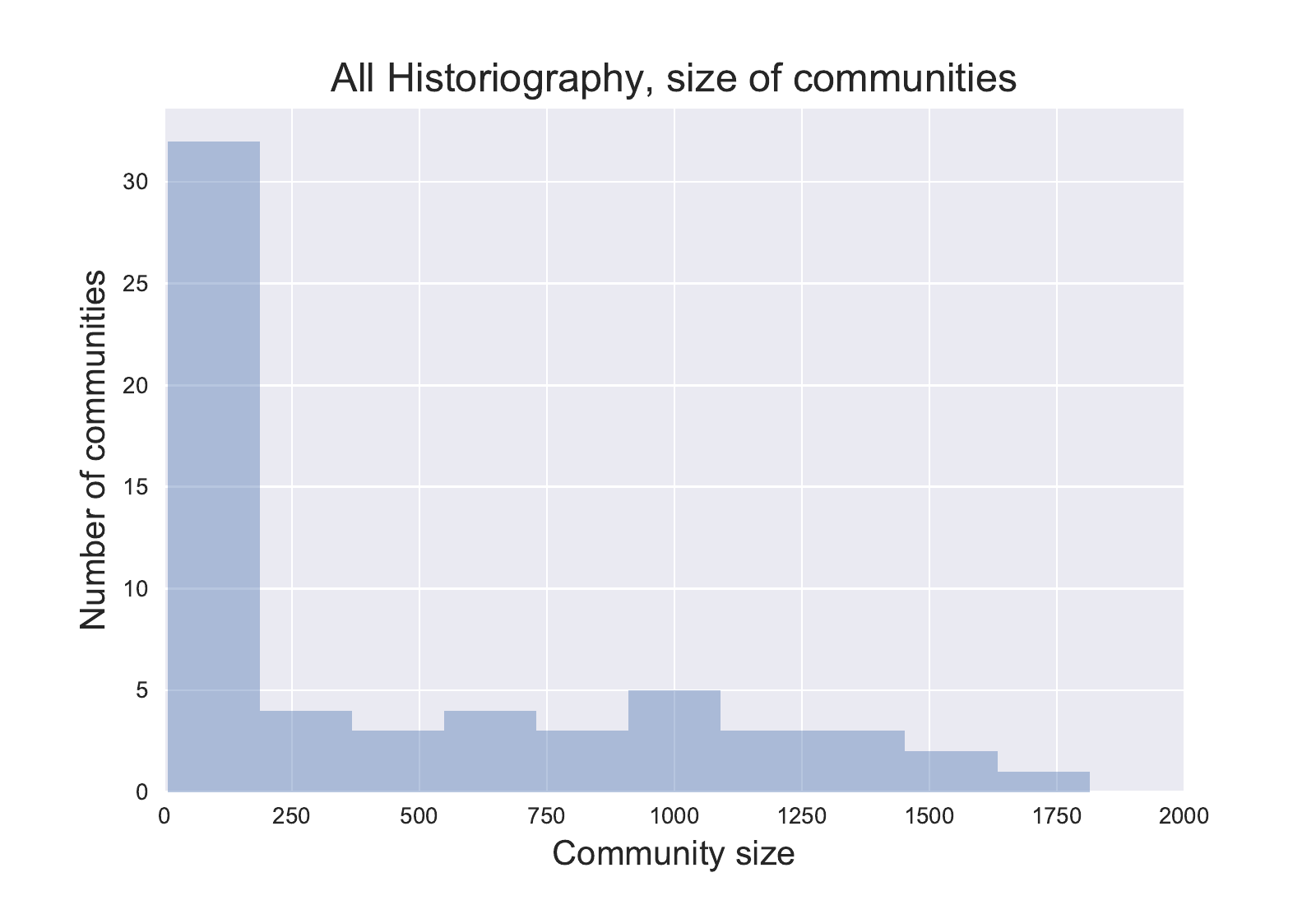}
	\caption{The distribution of the size of communities in the all history network, using the Louvain method with configuration identical to experiments. Shown are only communities with more than five articles. An inspection of the first six communities, by reading a random sample of 200 article titles each, led to the following broad classification (in order of decreasing size): Economic history; Intellectual and cultural history (pre-contemporary); Social history; Gender, slavery and minorities history; Colonial and post-colonial history; Contemporary political history.}
	\label{fig:size_all}
\end{figure}

The communities of this dataset are of more difficult evaluation, given the size of the network. Indeed, the size of these communities is also varied, as shown in Figure \ref{fig:size_all}. For these reasons, and given the low coverage of WoS with respect to history, results from this third dataset are included only for reference, in order to highlight the structural properties of a larger, more sparsely connected network.

\section{Results}
\label{sec:6}

The indicators applied on these datasets should allow us to highlight different properties of the core literature. We start with some hypotheses. First of all, that the core literature is present and plays a role into connecting the bibliographic coupling network of the respective dataset. Secondly, core books contribute more on average to the global connectivity of the network than core journal articles. If this were the case, their within indicator should be lower and their between indicator higher. Similarly, in such a case core books should be more likely to have a low bridging capacity, as they would not just connect two specific communities but several more. Lastly, we see no reason for their topicality to differ from that of journal articles, since it makes intuitive sense that any source is better known to a specific community than in general.

The starting point of the analysis are the bibliographic coupling networks of the three datasets. Table \ref{tab:bibc_stats} reports their summary statistics. Most notably, the three networks differ in the basic terms of their connectivity. The first dataset, monographs on the history of Venice, results in a very dense and well-connected network; less so for The Library; and quite less so for the history dataset from WoS, which comprises a 30\% of vertices in small components.

\begin{table}[H]
	\centering
	\caption{Summary statistics of the bibliographic coupling networks.}
	\label{tab:bibc_stats}
	\resizebox{\columnwidth}{!}{%
		\begin{tabular}{cccc}
			\textbf{Statistic/Dataset}               & \textbf{Monographs History Venice} & \textbf{The Library 1981-2016} & \textbf{All History 2005-2015} \\
			\textbf{\# vertices}                     & 673                                & 479                            & 36709                                 \\
			\textbf{\# edges}                        & 87168                              & 4435                           & 161802                                \\
			\textbf{\# connected components}         & 3                                  & 58                             & 9901                                  \\
			\textbf{Vertices in the giant component} & 99.7\%                             & 87.9\%                         & 69.5\%                                \\
			\textbf{Network density}                         & 0.38548                            & 0.03874                        & 0.00024                              
		\end{tabular}
	}
\end{table}

The statistics of indicators given in Table \ref{tab:indicators} confirm in part our initial hypotheses. The average modularity of partitions, influenced by the connectivity of the networks (higher connectivity usually implies a more difficult community partitioning task, thus lower resulting modularity), is much lower for the first dataset, and incrementally higher for the other two. 

\begin{table}[H]
	\centering
	\caption{Mean (median) value of the indicators over different datasets, plus the modularity of partitions. Values are averaged over ten partitions.}
	\label{tab:indicators}
	\resizebox{\columnwidth}{!}{%
		\begin{tabular}{cccc}
			\textbf{Statistic/Dataset}       & \textbf{Monographs History Venice} & \textbf{The Library 1981-2016} & \textbf{All History 2005-2015} \\
			\textbf{Within indicator}        & 0.18 (0.17)                        & 0.43 (0.43)                    & 0.78 (0.83)                           \\
			\textbf{Between indicator}       & -0.18 (-0.17)                      & -0.43 (-0.43)                  & -0.78 (-0.83)                         \\
			\textbf{Topicality indicator}    & 0.32 (0.37)                        & 0.41 (0.44)                    & 0.61 (0.62)                           \\
			\textbf{Bridging indicator}      & 0.16 (0.14)                        & 0.48 (0.48)                    & 0.45 (0.43)                           \\
			\textbf{Modularity of partition} & 0.1835                             & 0.4355                         & 0.7135                               
		\end{tabular}
	}
\end{table}

\begin{figure}[H]
	\begin{minipage}{0.32\textwidth}
		\centering\includegraphics[width=\textwidth]{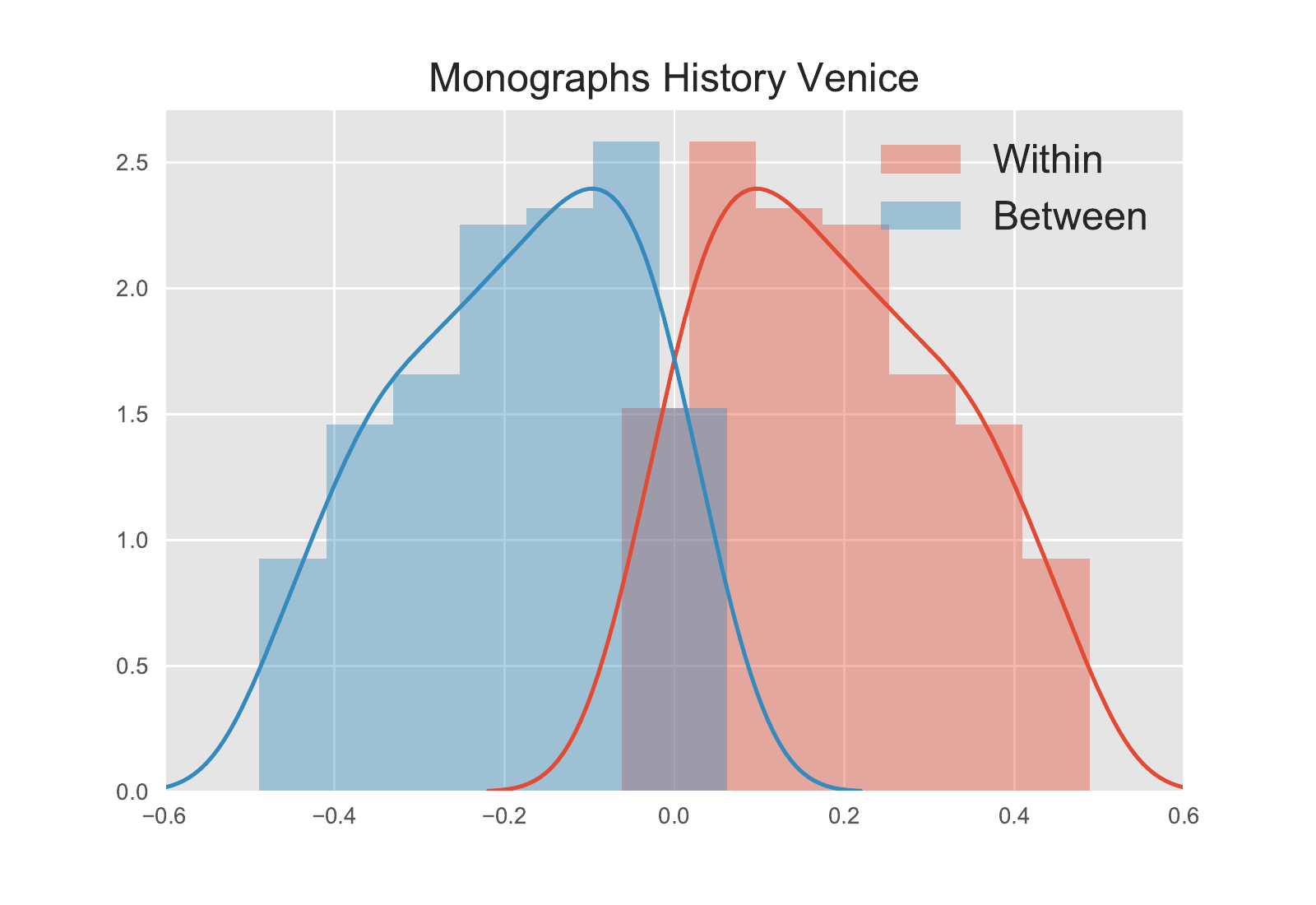}
		\subcaption{Monographs History Venice.}\label{fig:w_b_venice}
	\end{minipage}\hfill
	\begin{minipage}{0.32\textwidth}
		\centering\includegraphics[width=\textwidth]{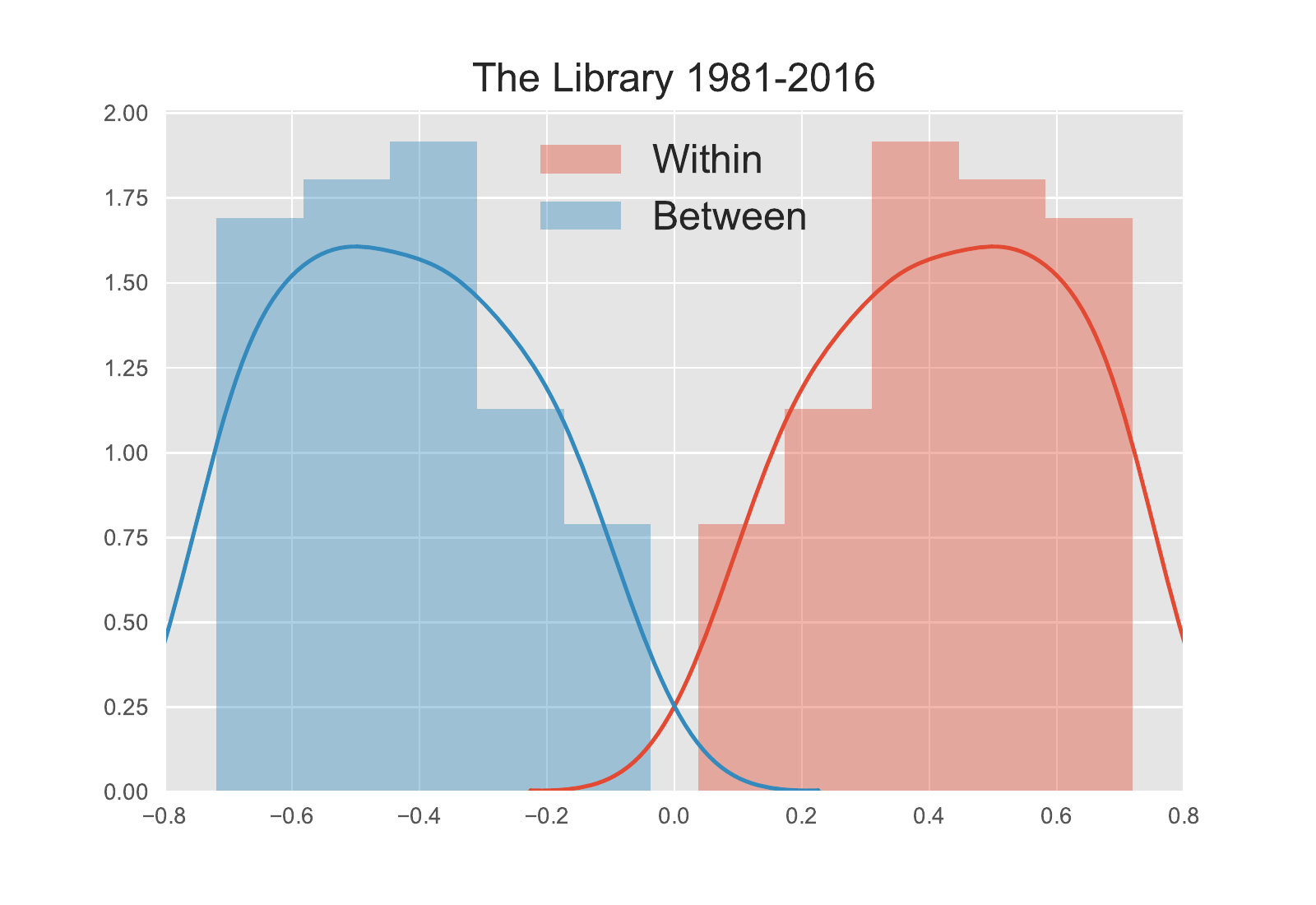}
		\subcaption{The Library \\1981-2016.}\label{fig:w_b_library}
	\end{minipage}\hfill
	\begin{minipage}{0.32\textwidth}
		\centering\includegraphics[width=\textwidth]{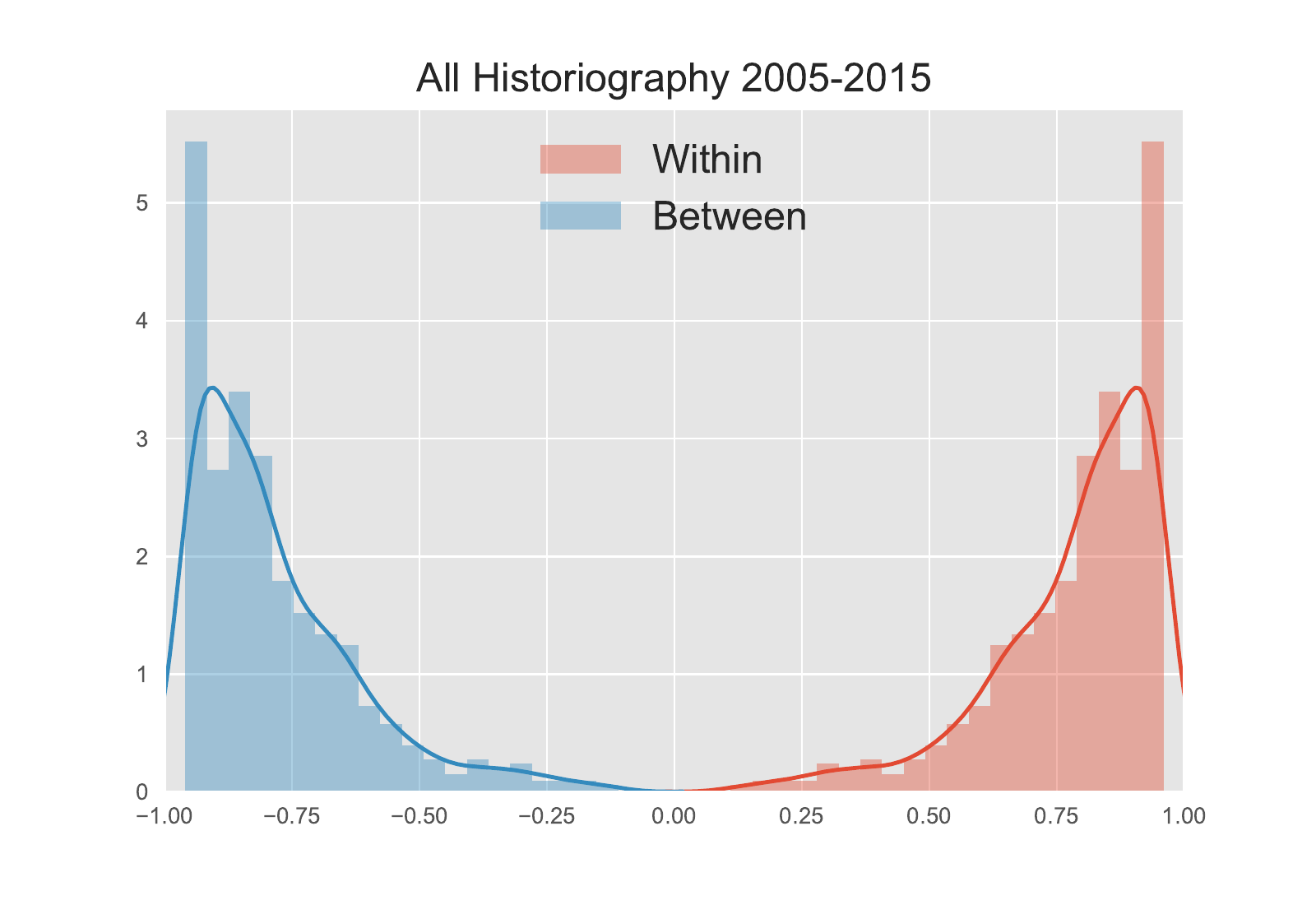}
		\subcaption{All History \\2005-2015.}\label{fig:w_b_all}
	\end{minipage}
	\caption{Distributions of the within and between indicators. Histograms are normalized, the lines are the kernel density estimations.}\label{fig:w_b}
\end{figure}

The within and between indicators behave as expected, their distributions are shown in Figure \ref{fig:w_b}. Indeed, a core literature made only of books acts not just within specific communities but across them. If we consider the third dataset, the action of the core is considerably more limited to providing within-community connectivity. The second dataset, where the core literature is mixed, lays in-between these two extremes.

\begin{figure}[H]
	\begin{minipage}{0.32\textwidth}
		\centering\includegraphics[width=\textwidth]{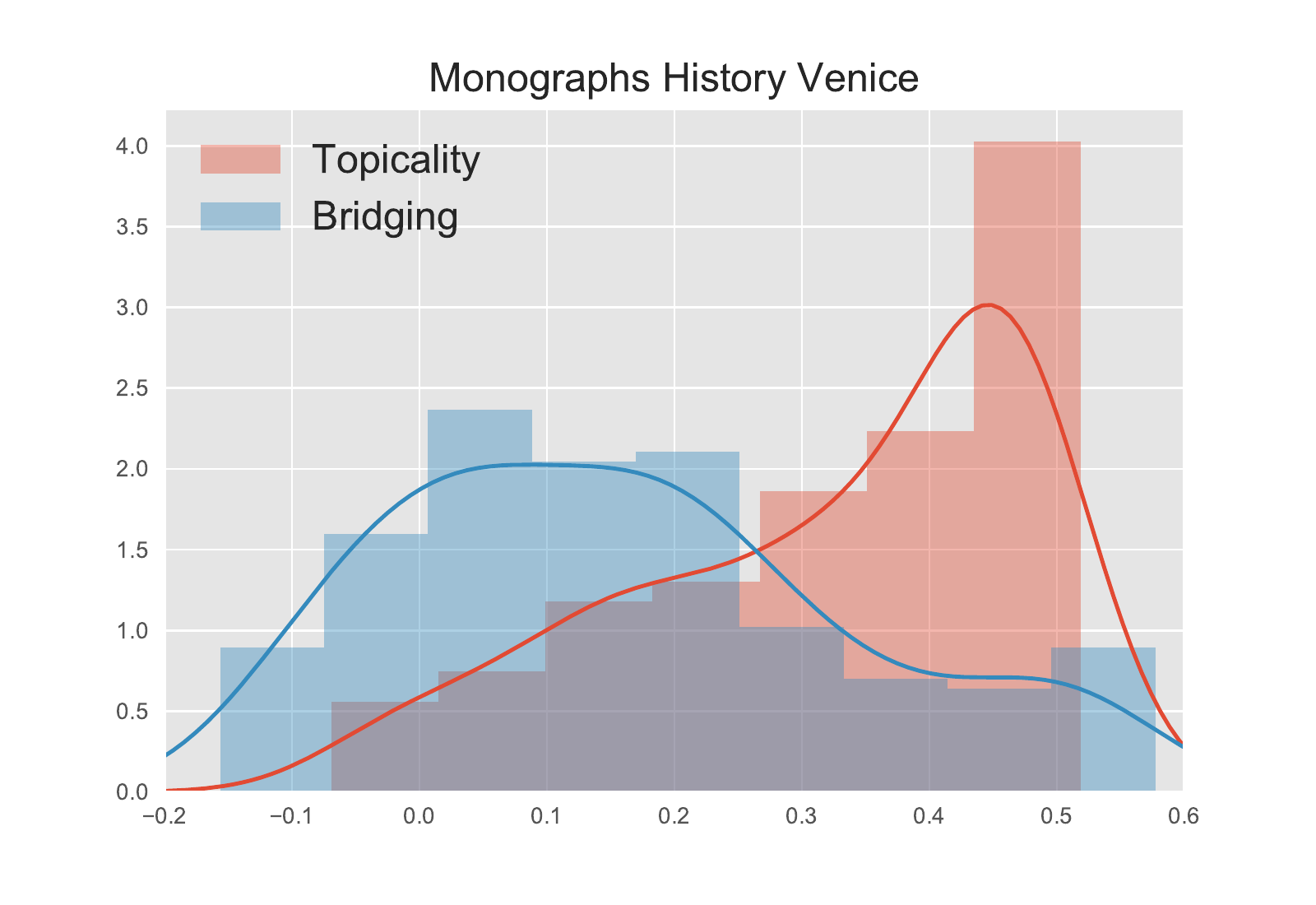}
		\subcaption{Monographs History Venice.}\label{fig:t_b_venice}
	\end{minipage}\hfill
	\begin{minipage}{0.32\textwidth}
		\centering\includegraphics[width=\textwidth]{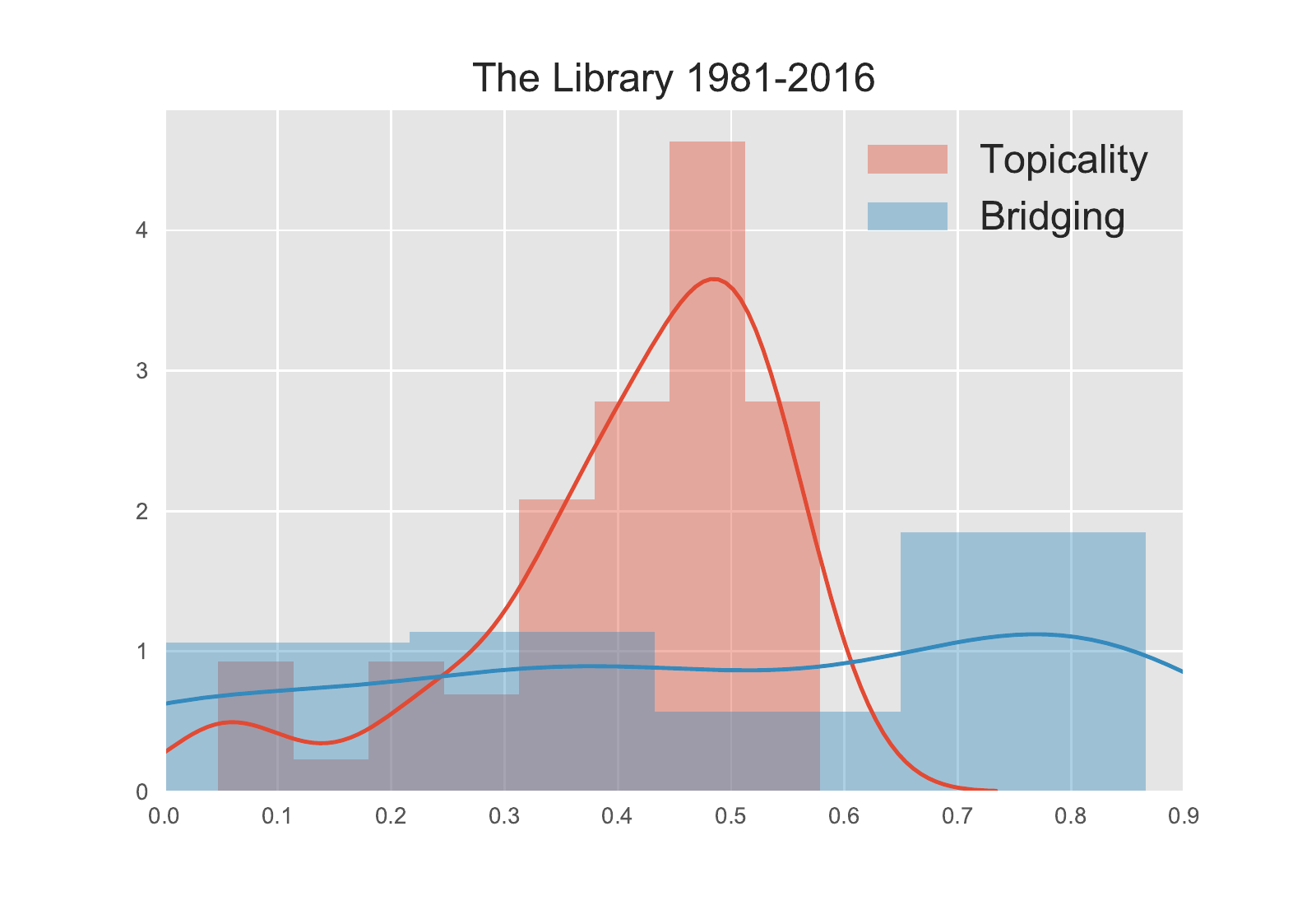}
		\subcaption{The Library \\1981-2016.}\label{fig:t_b_library}
	\end{minipage}\hfill
	\begin{minipage}{0.32\textwidth}
		\centering\includegraphics[width=\textwidth]{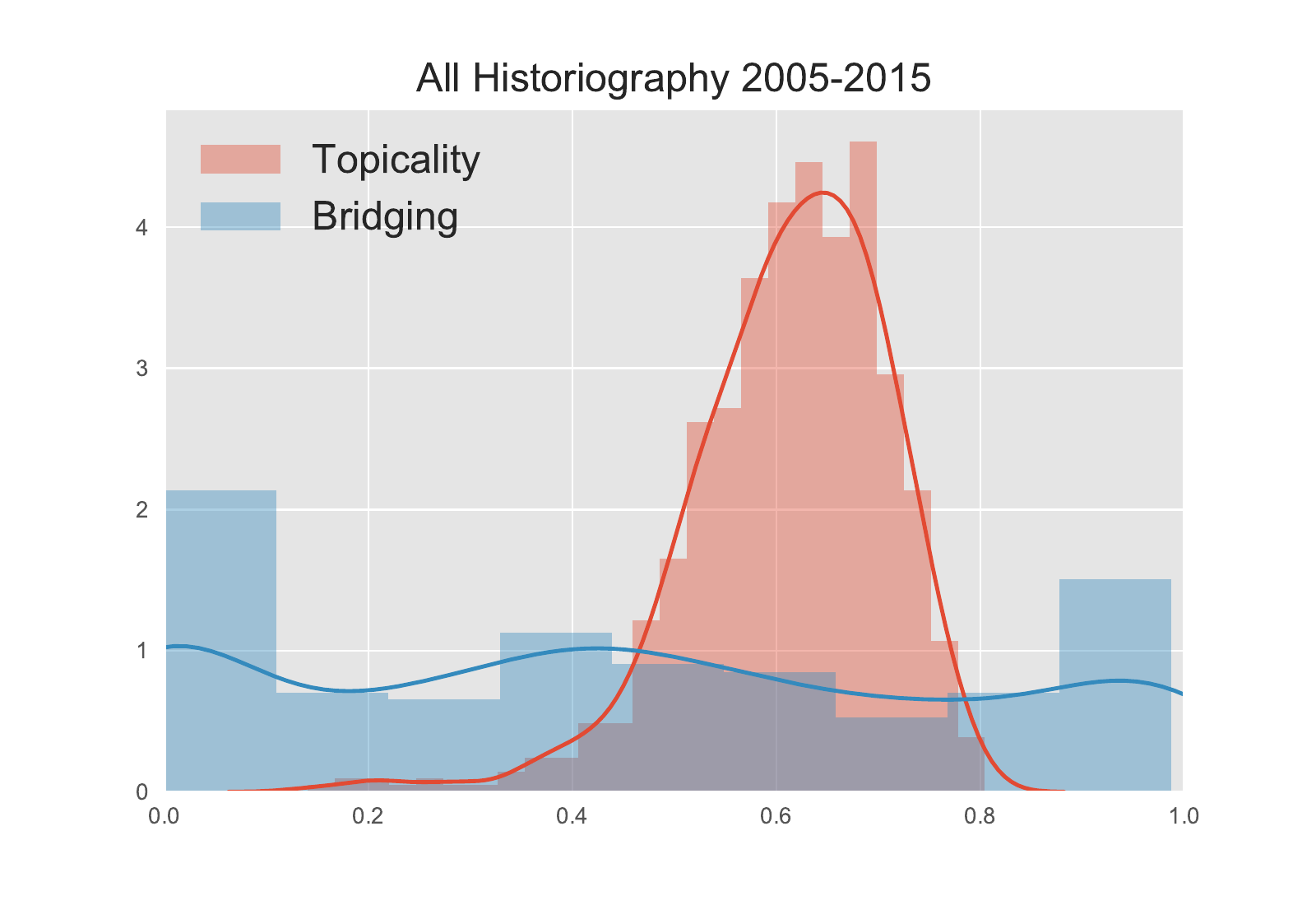}
		\subcaption{All History \\2005-2015.}\label{fig:t_b_all}
	\end{minipage}
	\caption{Distributions of the topicality and bridging indicators. Histograms are normalized, the lines are the kernel density estimations.}\label{fig:t_b}
\end{figure}

With respect to the topicality and bridging indicators, results are less unequivocal. Their distributions are given in Figure \ref{fig:t_b}. If we can appreciate that indeed topicality is increasingly higher moving from dataset one to three, bridging captures a variety of behaviours, resulting in an almost uniform distribution of values. Core sources can thus bridge several or few pairs of communities irrespective of their typology. Nevertheless, there is a clear higher concentration of low-value bridging core sources in the first dataset, as shown in Figure \ref{fig:t_b_venice}. The core literature from the monographs on Venice presents a higher proportion of low-bridging core sources, and a longer left tail in the topicality distribution as well. This entails that the connectivity action of these core sources is not just focused on a single community or community pair.

\begin{figure}[H]
	\begin{minipage}{0.32\textwidth}
		\centering\includegraphics[width=\textwidth]{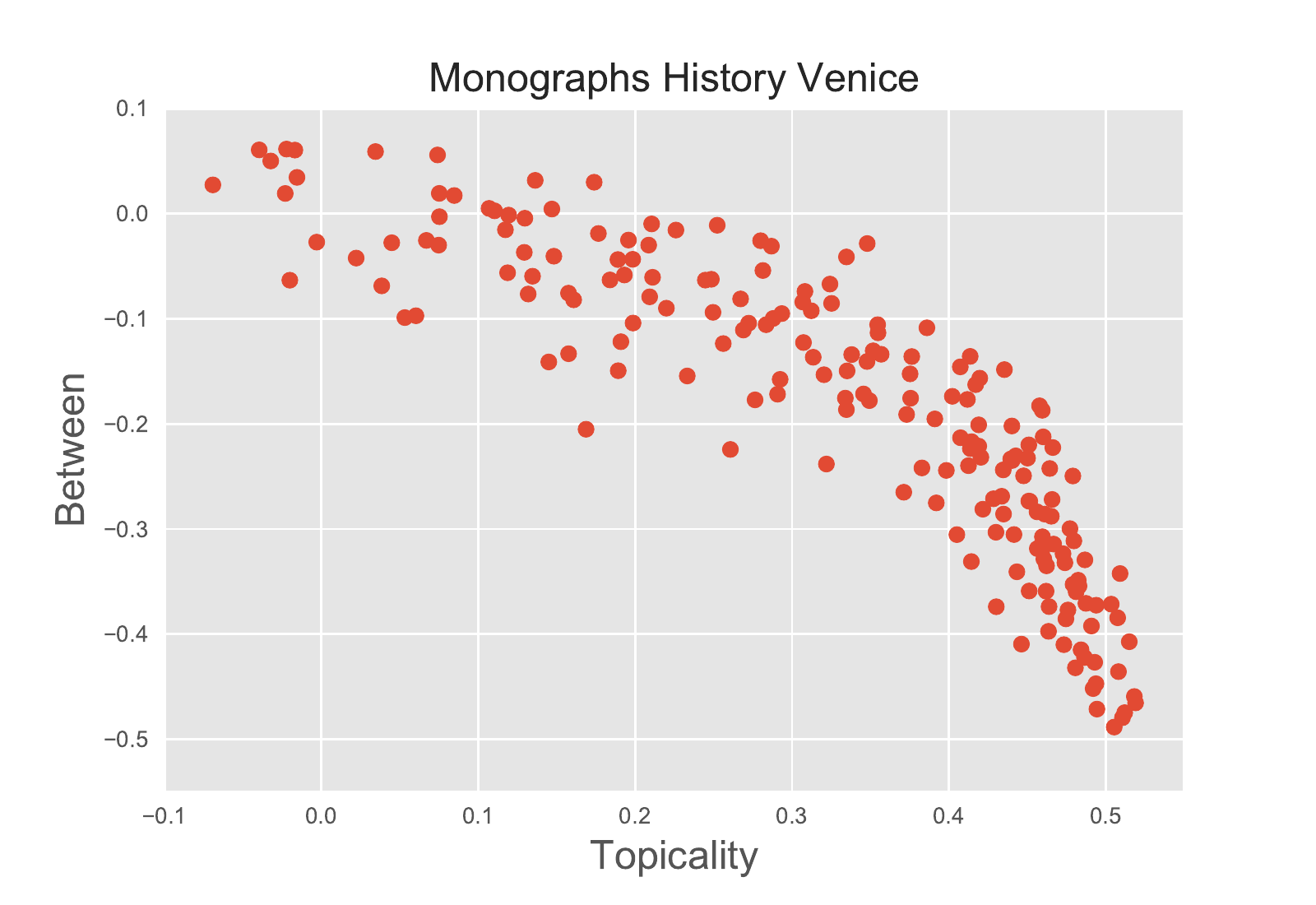}
		\subcaption{Monographs History Venice.}\label{fig:t_vs_b_venice}
	\end{minipage}\hfill
	\begin{minipage}{0.32\textwidth}
		\centering\includegraphics[width=\textwidth]{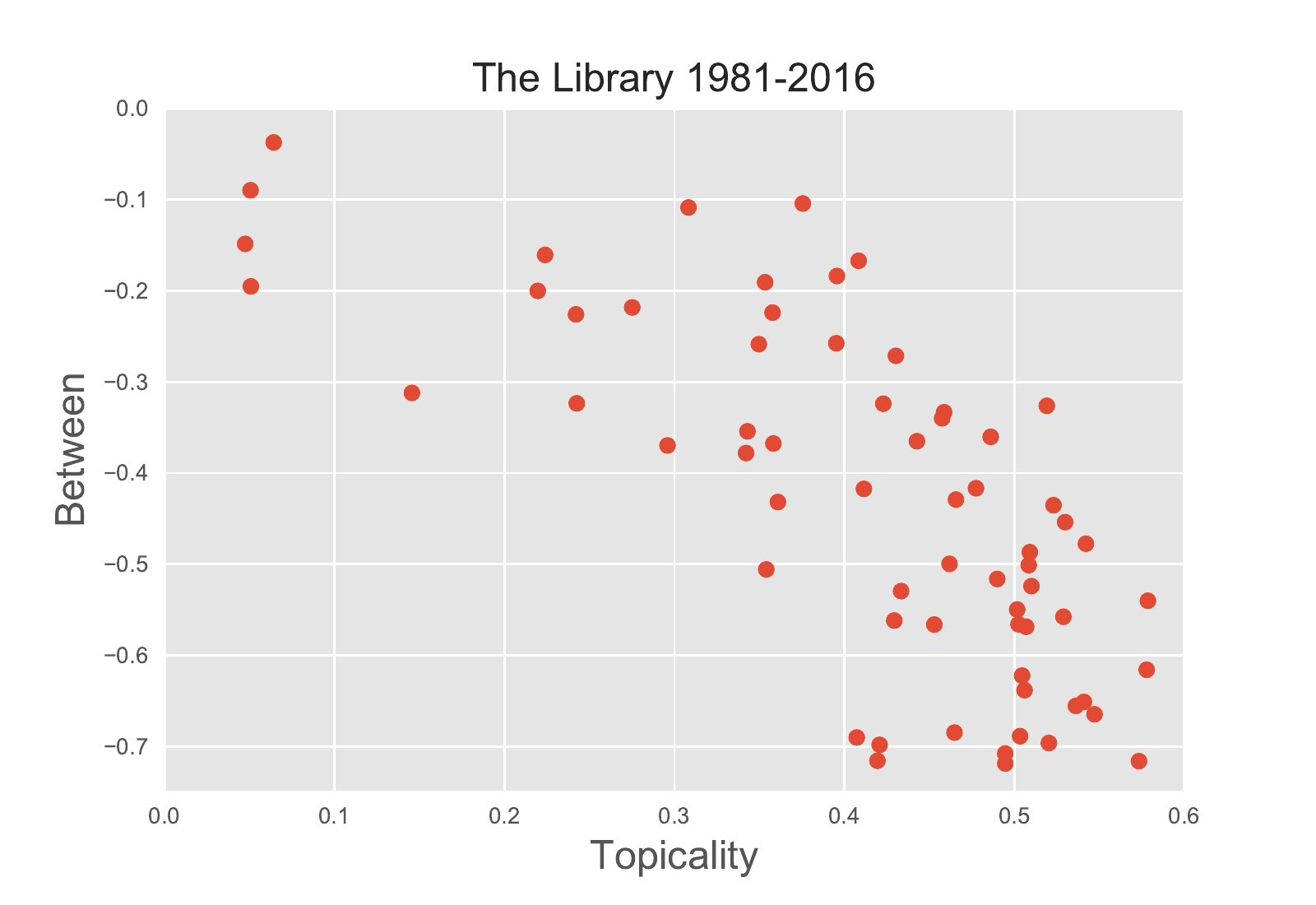}
		\subcaption{The Library \\1981-2016.}\label{fig:t_vs_b_library}
	\end{minipage}\hfill
	\begin{minipage}{0.32\textwidth}
		\centering\includegraphics[width=\textwidth]{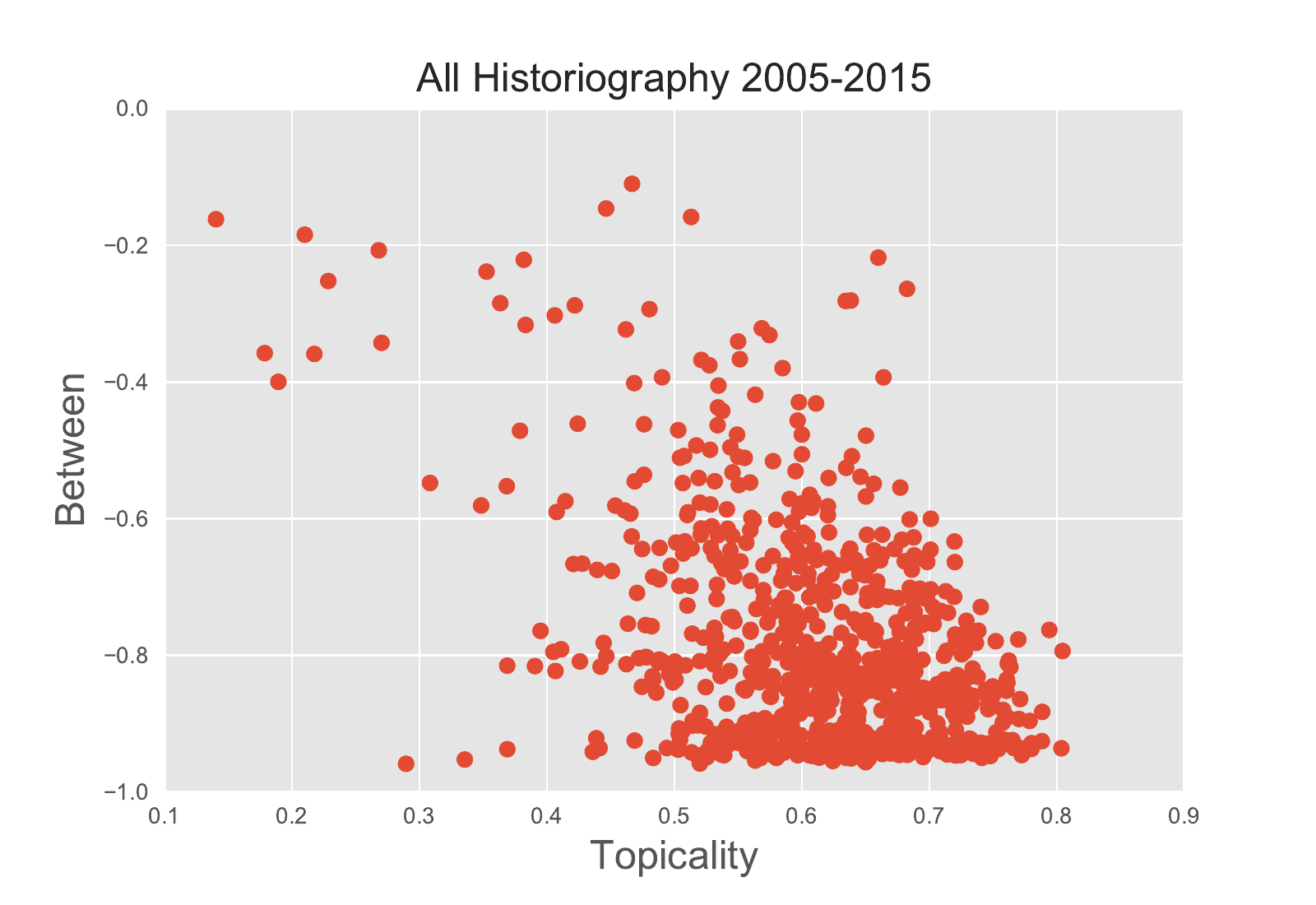}
		\subcaption{All History \\2005-2015.}\label{fig:t_vs_b_all}
	\end{minipage}
	\caption{Scatter plots of the topicality and between indicators. Note that the within indicator is symmetric (positive) of between, around zero.}\label{fig:t_vs_b}
\end{figure}

The more varied behaviour of a core literature composed of books is highlighted in scatter plots which consider topicality and between indicators at the same time, in Figure \ref{fig:t_vs_b}. In dataset one, a clear pattern exists by which low-topicality sources have high between score, and vice versa, in two distinct linear regimes of change. Conversely, the norm for sources in the third dataset is to have very high topicality and varied, but comparatively lower between scores. The second dataset, as usual, presents mixed results. 

The implication is relatively clear: a core literature composed of books contains a higher proportion of globally interconnecting sources, spanning way outside their main community. Such segment of the core literature interconnects communities at a global structural level, where the action of another part of the core literature, and most journal articles in it, is structurally localized.

\begin{comment}
It must be stressed again that dataset three should only be compared with an equally-sized dataset containing also book citations, in order to fairly account for its size. Communities in dataset three can count up to 2000 vertices, while for dataset one and two they reach 200 vertices maximum. This fact bears consequences with respect to indicators. Nevertheless, its results still show that even at a larger scale, core journal articles act mostly locally.
\end{comment}

\begin{figure}[H]
	\begin{minipage}{0.5\textwidth}
		\centering\includegraphics[width=\textwidth]{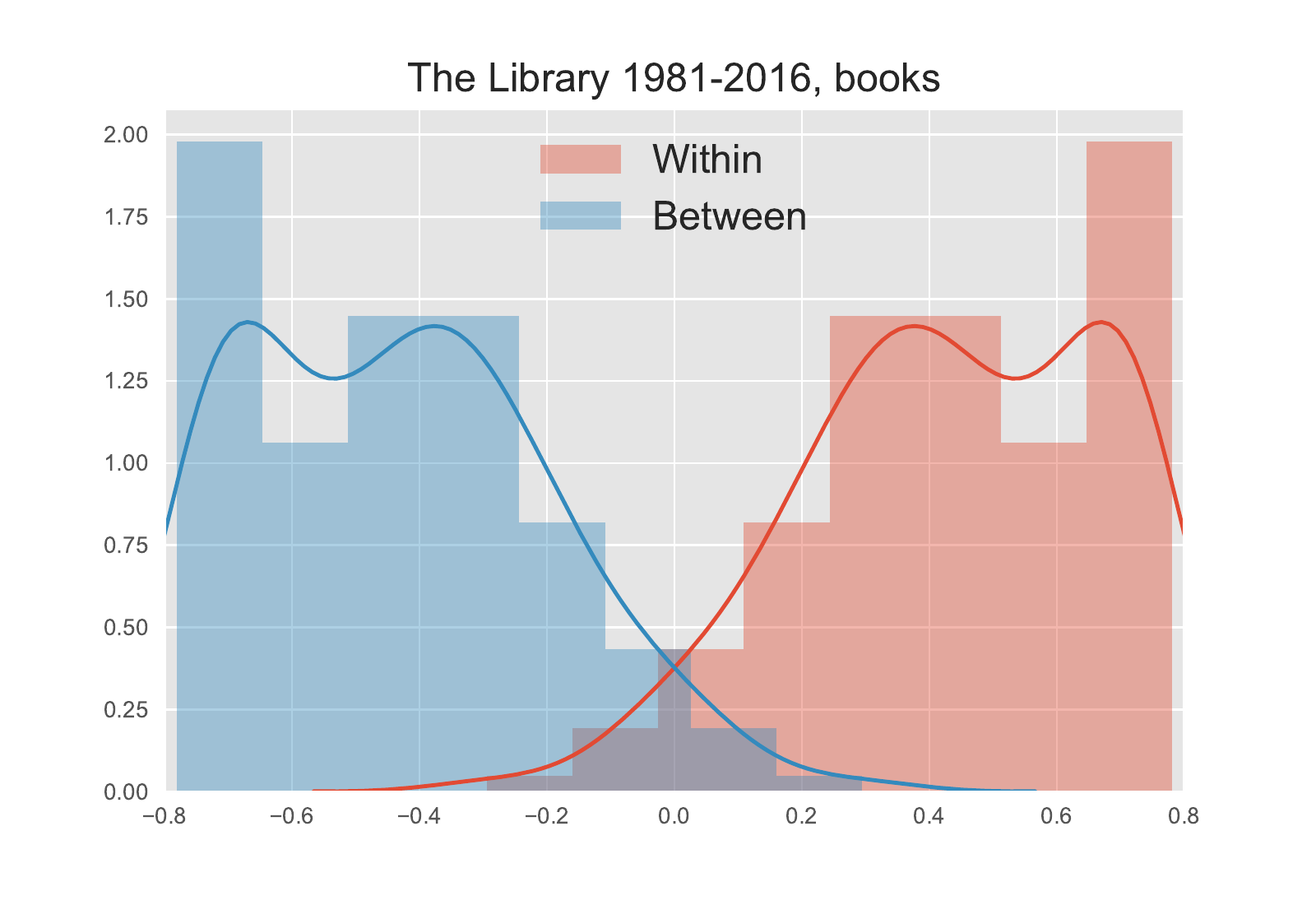}
		\subcaption{The Library \\1981-2016, books.}\label{fig:library_books}
	\end{minipage}\hfill
	\begin{minipage}{0.5\textwidth}
		\centering\includegraphics[width=\textwidth]{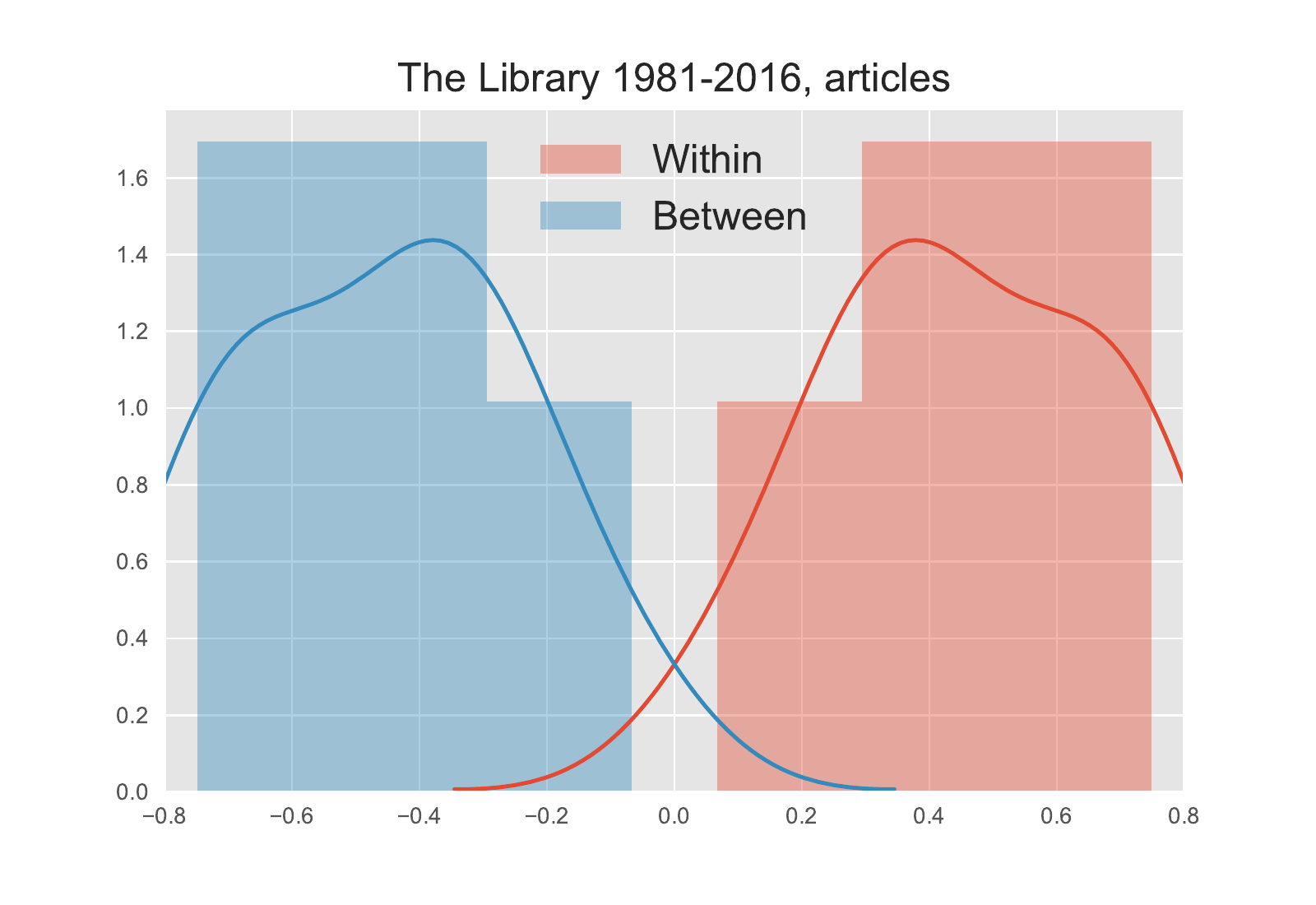}
		\subcaption{The Library \\1981-2016, journal articles.}\label{fig:library_articles}
	\end{minipage}
	\caption{Distributions of the within and between indicators for The Library, core books and journal articles respectively. Histograms are normalized, the lines are the kernel density estimations.}\label{fig:library_focus}
\end{figure}

To further explore the difference between core books and journal articles, we consider dataset two and enlarge the number of core sources under consideration by taking the 99 quantile. There are 167 core sources now, of which 154 are books and still only 13 journal articles. The distribution of their within and between indicators, given in Figure \ref{fig:library_focus}, highlights the higher global reach of some books, being also negative within score, not to be found among journal articles. To be sure, most books still act locally as do most journal articles.

The correlations among indicators further elucidate this preliminary distinction, as shown in Table \ref{tab:correlations}. For the first dataset, the within, topicality and bridging indicators act relatively in uniformity: they rise if the connectivity action of the core source is localized, they lower (and between indicator rises) if it is more global. In the second dataset this relation stands only for the within and topicality indicator, while bridging indicator loses correlation. For the last dataset, the relation lowers even more, at times becoming negative, and the main positive correlation is now between the within indicator and in-degree values. These results confirm the existence of two distinct ways of `being core': with a global or localized connectivity action. In the case of the first dataset, a local action can happen: within communities, and usually within a specific one, or bridging a given pair of communities. This was also clear from Figure \ref{fig:t_vs_b_venice}, as the between indicator is symmetric around zero to the within one, topicality is clearly positively correlated with the within indicator. A global action entails instead acting between several of these community pairs. For the last dataset, the action of the core (in this case, journal articles), is essentially only local either to a given community or to several communities, with few exceptions.

\begin{table}[H]
	\centering
	\caption{Correlations of the indicators and the in-degree of the core sources. The between indicator is omitted as superfluous. Pearson: top-right. Spearman: bottom-left.}
	\label{tab:correlations}
	\resizebox{\columnwidth}{!}{%
		\begin{tabular}{cccccc}
			\textbf{Dataset}                                                            & \textbf{Indicator}                       & \textbf{Within}            & \textbf{Topicality}       & \textbf{Bridging}          & \textbf{Indegree} \\ \hline
			\multicolumn{1}{c|}{\multirow{4}{*}{\textbf{Monographs History Venice}}}    & \multicolumn{1}{c|}{\textbf{Within}}     & \multicolumn{1}{c|}{1}     & \multicolumn{1}{c|}{0.86} & \multicolumn{1}{c|}{0.62}  & -0.13             \\
			\multicolumn{1}{c|}{}                                                       & \multicolumn{1}{c|}{\textbf{Topicality}} & \multicolumn{1}{c|}{0.92}  & \multicolumn{1}{c|}{1}    & \multicolumn{1}{c|}{0.38}  & -0.08             \\
			\multicolumn{1}{c|}{}                                                       & \multicolumn{1}{c|}{\textbf{Bridging}}   & \multicolumn{1}{c|}{0.63}  & \multicolumn{1}{c|}{0.37} & \multicolumn{1}{c|}{1}     & -0.19             \\
			\multicolumn{1}{c|}{}                                                       & \multicolumn{1}{c|}{\textbf{Indegree}}   & \multicolumn{1}{c|}{-0.03} & \multicolumn{1}{c|}{0.02} & \multicolumn{1}{c|}{-0.14} & 1                 \\ \hline
			\multicolumn{1}{c|}{\multirow{4}{*}{\textbf{The Library 1981-2016}}}        & \multicolumn{1}{c|}{\textbf{Within}}     & \multicolumn{1}{c|}{1}     & \multicolumn{1}{c|}{0.72} & \multicolumn{1}{c|}{-0.01} & 0.05              \\
			\multicolumn{1}{c|}{}                                                       & \multicolumn{1}{c|}{\textbf{Topicality}} & \multicolumn{1}{c|}{0.7}   & \multicolumn{1}{c|}{1}    & \multicolumn{1}{c|}{0.08}  & 0.06              \\
			\multicolumn{1}{c|}{}                                                       & \multicolumn{1}{c|}{\textbf{Bridging}}   & \multicolumn{1}{c|}{0.01}  & \multicolumn{1}{c|}{0.09} & \multicolumn{1}{c|}{1}     & 0.11              \\
			\multicolumn{1}{c|}{}                                                       & \multicolumn{1}{c|}{\textbf{Indegree}}   & \multicolumn{1}{c|}{0}     & \multicolumn{1}{c|}{0.2}  & \multicolumn{1}{c|}{0.09}  & 1                 \\ \hline
			\multicolumn{1}{c|}{\multirow{4}{*}{\textbf{All History 2005-2015}}} & \multicolumn{1}{c|}{\textbf{Within}}     & \multicolumn{1}{c|}{1}     & \multicolumn{1}{c|}{0.44} & \multicolumn{1}{c|}{-0.15} & 0.18              \\
			\multicolumn{1}{c|}{}                                                       & \multicolumn{1}{c|}{\textbf{Topicality}} & \multicolumn{1}{c|}{0.31}  & \multicolumn{1}{c|}{1}    & \multicolumn{1}{c|}{-0.01} & 0.44              \\
			\multicolumn{1}{c|}{}                                                       & \multicolumn{1}{c|}{\textbf{Bridging}}   & \multicolumn{1}{c|}{-0.42} & \multicolumn{1}{c|}{0.01} & \multicolumn{1}{c|}{1}     & 0.05              \\
			\multicolumn{1}{c|}{}                                                       & \multicolumn{1}{c|}{\textbf{Indegree}}   & \multicolumn{1}{c|}{0.12}  & \multicolumn{1}{c|}{0.57} & \multicolumn{1}{c|}{0.08}  & 1                
		\end{tabular}
	}
\end{table}

By taking a look at the top core sources for each dataset, according to every indicator, we can get a more concrete idea of the different structural roles core sources can have. Lists are provided in the Appendix: Tables \ref{tab:top_venice}, \ref{tab:top_library} and \ref{tab:top_all} for datasets one, two and three respectively. In general, the top of the between indicator indeed captures the sources most transversal to several scholarly communities. In the case of the first and second datasets, these are usually cornerstone monographs or reference works, in the last dataset we find instead methodological papers.

Other indicators follow instead the behaviour previously discussed. For example, the top sources by within and topicality for the first dataset on Venice, contain sources specific to the art and architecture history of the city.

Our results thus point to the presence of a core literature which is, where applicable, mostly composed of books. Book citations indeed seem to considerably rise the connectivity of the bibliographic coupling network. We also highlighted at least two structural actions that the core literature can play: a \textit{local action}, connecting within one or more communities, and a \textit{global action}, connecting across communities. The global one is mainly performed by some monographs and reference works, more rarely by journal articles. As it stands now, we are left wondering if the core literature is also dependent on the citing publications. Despite this remaining an open question, it would not appear so. Our second dataset, considering article citations to both WoS source and non-source items, effectively shows a mix behaviour, in-between the first and third datasets.

\section{Conclusions}
\label{sec:7}

The accumulation of knowledge is a topic of great interest in bibliometrics. In the humanities, such is the variety of publication venues, typologies and languages, that it is difficult to disentangle the effects of each component of this \textit{multidimensional system of knowledge}. We interested ourselves here with one discipline, history, and with one element of the system, the core literature. We defined it as the most cited sources of a given representative dataset, and asked the following questions: a) is there a well-defined core literature in history at different publication levels? b) What structural role does it play in the definition of the intellectual landscape of citing publications?

In order to answer such questions, we considered the bibliographic coupling network of citing publications, and their partitions into communities. Four indicators were introduced: a) the \textit{within indicator} maps the action of a core source into connecting vertices within communities; b) the \textit{between indicator} maps this action between pairs of communities; c) the \textit{topicality indicator} assesses the proportion of the within action happening in a unique community; d) the \textit{bridging indicator} assessed the proportion of the between action happening in a unique pair of communities. All indicators account for the skewed effect of the in-degree of the core literature by filtering it out using a null configuration model. The proposed method can be used in general to investigate the structural contribution of vertices after network projections such as bibliographic coupling. Three datasets were considered in order to explore the role of core literatures from different perspectives: a dataset of monograph to book citations, a dataset of journal articles to all source and non-source WoS items they cite, and a dataset including all journal article to journal article citations in history, as indexed in WoS over eleven years (2005-2015). 

The main finding is that the core literature, clearly emerging in all datasets under consideration, has at least two distinct structural effects on the intellectual landscape of the citing publications. This effect can either be \textit{localized}, by rising the connectivity within one or several communities, or at times a specific pair of communities, or it can be \textit{global}, by rising the overall connectivity of the landscape, across communities. We also found that global action is usually performed by core sources which are well-known scholarly monographs, works of reference or primary sources. Local action can instead by performed by all kinds of core sources, and especially so by journal articles. As a result, the intellectual landscape becomes better connected by considering citations to books, which at times span more broadly, and not just journal articles, which usually remain known within specific scholarly communities. Indeed, where applicable, the core literature is mostly composed of books. These results will need to be complemented by an investigation of different specialisms in history and disciplines in the humanities, and by considering further cited sources, especially so unpublished primary sources. Future work should also explore the effective role of the core literature for a scholarly community: is it perfunctory or still intellectually relevant?

\begin{comment}

Our results allow for a few more general considerations. The intellectual landscapes of the humanities are complex items to study, which require considering the full range of objects at the cited side of the citation relation. The citing side might also be varied, with different behaviours according to the publication typology under consideration. This remains an open question. Secondly, the presence and structural action of the core literature can be used to help answering the information needs of scholars. Last but not least, the importance of considering all cited sources when evaluating the humanities cannot be stressed enough. For example, some of the most enduring contributions in historiography, and the humanities alike, are resources, works of reference, editions of primary sources, databases (also in their pre-digital forms). Their influence, as that of particularly important scholarly monographs, stands the test of the ages and is difficult to appreciate just by considering narrow time windows. Yet their impact can hardly be overestimated, given the enduring importance such works evidently have over decades and sometimes centuries.

\end{comment}

\section*{Acknowledgements}
	I would like to thank, in alphabetical order: Martina Babetto, Silvia Ferronato and Matteo Romanello (EPFL), colleagues who helped with the collection and processing of the Venetian dataset. The Library of the \textit{Ca' Foscari} University of Venice willingly collaborated with bibliographical resources and logistics support. The Central Institute for the Union Catalogue of Italian Libraries and Bibliographic Information (ICCU) shared its catalog metadata with us. I thank both for their support. I would also like to thank Ludo Waltman and Vincent Traag (CWTS Leiden), and Massimo Franceschet (University of Udine) for very helpful advice. I worked in part on this research project during a research stay at the Centre for Science and Technology Studies (CWTS) at Leiden University, where I had the opportunity to present and discuss my work with great benefits. I would finally like to thank CWTS for providing access to its databases.

\bibliographystyle{abbrvnat}
\bibliography{bib.bib}   % name your BibTeX data base

\section*{Appendix}
\label{sec:appendix}

% \usepackage{multirow}
\begin{table}[H]
	\centering
	\caption{Top-five core sources per indicator: Monographs History Venice.}
	\label{tab:top_venice}
	\resizebox{\columnwidth}{!}{%
		\begin{tabular}{|c|c|l|l|l|}
			\hline
			\textbf{Indicator}               & \textbf{Value} & \textbf{Author} & \textbf{Year} & \textbf{Title}                                                             \\ \hline
			\multirow{5}{*}{\textbf{Within}}     & 0.49           & Zanetti, A.M.   & 1733          & Descrizione di tutte le pubbliche pitture della città di Venezia           \\ \cline{2-5} 
			& 0.48           & Temanza, T.     & 1778          & Vite dei più celebri architetti e scultori veneziani                       \\ \cline{2-5} 
			& 0.47           & Zanotto, F.     & 1856          & Nuovissima guida di Venezia                                                \\ \cline{2-5} 
			& 0.47           & Aymard, M.      & 1966          & Venise, Raguse et le commerce du blé                                       \\ \cline{2-5} 
			& 0.46           & Scamozzi, V.    & 1615          & L'idea dell'architettura universale                                        \\ \hline
			\multirow{5}{*}{\textbf{Between}}    & 0.06           & Filiasi, G.     & 1811          & Memorie storiche de' Veneti                                                \\ \cline{2-5} 
			& 0.06           & Monticolo, G.   & 1896          & I capitolari delle Arti veneziane                                          \\ \cline{2-5} 
			& 0.06           & Soranzo, G.     & 1895          & Bibliografia veneziana                                                     \\ \cline{2-5} 
			& 0.06           & Canal, M.       & 1972          & Les estoires de Venise                                                     \\ \cline{2-5} 
			& 0.06           & Molmenti, P.    & 1973          & La storia di Venezia nella vita privata                                    \\ \hline
			\multirow{5}{*}{\textbf{Topicality}} & 0.52           & Scamozzi, V.    & 1615          & L'idea dell'architettura universale                                        \\ \cline{2-5} 
			& 0.52           & Zanetti, A.M.   & 1771          & Della pittura veneziana                                                    \\ \cline{2-5} 
			& 0.51           & Ridolfi, C.     & 1914          & Le maraviglie dell'arte                                                    \\ \cline{2-5} 
			& 0.51           & Zanotto, F.     & 1856          & Nuovissima guida di Venezia                                                \\ \cline{2-5} 
			& 0.51           & Temanza, T.     & 1778          & Vite dei più celebri architetti e scultori veneziani                       \\ \hline
			\multirow{5}{*}{\textbf{Bridging}}   & 0.58           & Moschini, G.    & 1815          & Guida per la città di Venezia                                              \\ \cline{2-5} 
			& 0.55           & Borsari, S.     & 1963          & Il dominio veneziano a Creta                                               \\ \cline{2-5} 
			& 0.54           & Preto, P.       & 1975          & Venezia e i Turchi                                                         \\ \cline{2-5} 
			& 0.54           & Rösch, G.       & 1989          & Der venezianische Adel bis zur Schliessung des Grossen Rats                \\ \cline{2-5} 
			& 0.54           & Pensolli, L.    & 1970          & La gerarchia delle fonti di diritto nella legislazione medievale veneziana \\ \hline
		\end{tabular}
	}
\end{table}

% \usepackage{multirow}
\begin{table}[H]
	\centering
	\caption{Top-five core sources per indicator: The Library 1981-2016.}
	\label{tab:top_library}
	\resizebox{\columnwidth}{!}{%
		\begin{tabular}{|c|c|l|l|l|}
			\hline
			\textbf{Indicator}               & \textbf{Value} & \textbf{Author} & \textbf{Year} & \textbf{Title}                                        \\ \hline
			\multirow{5}{*}{\textbf{Within}}     & 0.72           & Blagden, C.     & 1977          & The Stationers' Company: A History, 1403-1959         \\ \cline{2-5} 
			& 0.72           & Plomer, H.R.    & 1925          & Wynkyn de Worde \& His Contemporaries                 \\ \cline{2-5} 
			& 0.72           & Ker, N.R.       & 1987          & Medieval libraries of Great Britain                   \\ \cline{2-5} 
			& 0.71           & Shakespeare, W. & 1606          & King Lear                                             \\ \cline{2-5} 
			& 0.70           & Brusendorff, A. & 1925          & The Chaucer Tradition                                 \\ \hline
			\multirow{5}{*}{\textbf{Between}}    & -0.04          & Carter, H.      & 1975          & The Oxford University Press                           \\ \cline{2-5} 
			& -0.09          & Venn, J.A.      & 1951          & Alumni Cantabrigienses                                \\ \cline{2-5} 
			& -0.1           & McKerrow, R.B.  & 1927          & An Introduction to Bibliography for Literary Students \\ \cline{2-5} 
			& -0.11          & Stow, J.        & 1908          & A survey of London                                    \\ \cline{2-5} 
			& -0.15          & McKenzie, D.F.  & 1978          & Stationers' Company Apprentices                       \\ \hline
			\multirow{5}{*}{\textbf{Topicality}} & 0.58           & Oldham, J.B.    & 1952          & English Blind-Stamped Bindings                        \\ \cline{2-5} 
			& 0.58           & Greg, W.W.      & 1967          & A Companion to Arber                                  \\ \cline{2-5} 
			& 0.57           & Plomer, H.R.    & 1925          & Wynkyn de Worde \& His Contemporaries                 \\ \cline{2-5} 
			& 0.55           & Hodnett, E.     & 1935          & English Woodcuts 1480-1535                            \\ \cline{2-5} 
			& 0.54           & Maxted, I.      & 1977          & The London book trades, 1775-1800                     \\ \hline
			\multirow{5}{*}{\textbf{Bridging}}   & 0.85           & Smith, J.       & 1755          & Printer's Grammar                                     \\ \cline{2-5} 
			& 0.85           & Arber, E.       & 1903+         & The Term Catalogues                                   \\ \cline{2-5} 
			& 0.85           & Morrison, P.G.  & 1955          & Index of Printers, Publishers and Booksellers         \\ \cline{2-5} 
			& 0.84           & Foxon, D.F.     & 1975          & English Verse, 1701-1750                              \\ \cline{2-5} 
			& 0.84           & Lowry, M.       & 1979          & The World of Aldus Manutius                           \\ \hline
		\end{tabular}
	}
\end{table}

% \usepackage{multirow}
\begin{table}[H]
	\centering
	\caption{Top-five core sources per indicator: All History 2005-2015.}
	\label{tab:top_all}
	\resizebox{\columnwidth}{!}{%
		\begin{tabular}{|c|c|l|l|l|}
			\hline
			\textbf{Indicator}               & \textbf{Value} & \textbf{Author}  & \textbf{Year} & \textbf{Title}                                                                                                                                                                                      \\ \hline
			\multirow{5}{*}{\textbf{Within}}     & 0.96           & Karr, R.D.       & 1998          & \begin{tabular}[c]{@{}l@{}}``Why should you be so furious?": \\ The violence of the Pequot War\end{tabular}                                                                                          \\ \cline{2-5} 
			& 0.96           & Williams, S.     & 2005          & \begin{tabular}[c]{@{}l@{}}Poor relief, labourers' households and \\ living standards in rural England c. 1770-1834: \\ a Bedfordshire case study\end{tabular}                                      \\ \cline{2-5} 
			& 0.96           & Holquist, P.     & 2010          & \begin{tabular}[c]{@{}l@{}}``In Accord with State Interests and the People's Wishes": \\ The Technocratic Ideology of Imperial Russia's \\ Resettlement Administration\end{tabular}                  \\ \cline{2-5} 
			& 0.95           & Krige, J.        & 2006          & Atoms for peace, scientific internationalism, and scientific intelligence                                                                                                                           \\ \cline{2-5} 
			& 0.95           & Runia, E.        & 2007          & Burying the dead, creating the past                                                                                                                                                                 \\ \hline
			\multirow{5}{*}{\textbf{Between}}    & -0.11          & Aslanian, S.D.   & 2013          & \begin{tabular}[c]{@{}l@{}}AHR Conversation How Size Matters: \\ The Question of Scale in History\end{tabular}                                                                                      \\ \cline{2-5} 
			& -0.15          & Harvey, D.       & 1990          & Between Space and Time: Reflections on the Geographical Imagination                                                                                                                                 \\ \cline{2-5} 
			& -0.16          & Stoler, A.L.     & 2006          & On Degrees of Imperial Sovereignty                                                                                                                                                                  \\ \cline{2-5} 
			& -0.16          & White, H.        & 1984          & The Question of Narrative in Contemporary Historical Theory                                                                                                                                         \\ \cline{2-5} 
			& -0.18          & Mann, G.         & 2005          & \begin{tabular}[c]{@{}l@{}}Locating Colonial Histories: \\ Between France and West Africa\end{tabular}                                                                                              \\ \hline
			\multirow{5}{*}{\textbf{Topicality}} & 0.8            & Bradley, J.      & 2002          & \begin{tabular}[c]{@{}l@{}}Subjects into citizens: \\ Societies, civil society, and autocracy in tsarist Russia\end{tabular}                                                                        \\ \cline{2-5} 
			& 0.8            & Nora, P.         & 1989          & Between memory and history, les lieux-de-memoire                                                                                                                                                    \\ \cline{2-5} 
			& 0.79           & Weitz, E.D.      & 2008          & \begin{tabular}[c]{@{}l@{}}From the Vienna to the Paris System: \\ International Politics and the Entangled Histories of Human Rights, \\ Forced Deportations, and Civilizing Missions\end{tabular} \\ \cline{2-5} 
			& 0.79           & Spear, T.        & 2003          & Neo-traditionalism and the limits of invention in British colonial Africa                                                                                                                           \\ \cline{2-5} 
			& 0.79           & Werner, M.       & 2006          & Beyond comparison: Histoire croisee and the challenge of reflexivity                                                                                                                                \\ \hline
			\multirow{5}{*}{\textbf{Bridging}}   & 0.99           & Scott, J.W.      & 1991          & The evidence of experience                                                                                                                                                                          \\ \cline{2-5} 
			& 0.99           & Foucault, M.     & 1986          & Of other spaces                                                                                                                                                                                     \\ \cline{2-5} 
			& 0.99           & Huntington, S.P. & 1993          & The clash of civilizations                                                                                                                                                                          \\ \cline{2-5} 
			& 0.99           & Spear, T.        & 2003          & Neo-traditionalism and the limits of invention in British colonial Africa                                                                                                                           \\ \cline{2-5} 
			& 0.99           & Greif, A.        & 1993          & \begin{tabular}[c]{@{}l@{}}Contract enforceability and economic institutions in early trade: \\ the Maghreb traders coalition\end{tabular}                                                          \\ \hline
		\end{tabular}
	}
\end{table}

\end{document}